\documentclass[12pt]{article}
\usepackage{amsmath,amssymb}
\usepackage{graphicx}

\newcommand{\eqdef}{\stackrel{\text{def}}{=}}
\newcommand{\n}{\nonumber \\}
\newcommand{\bm}{\boldsymbol}
\newcommand{\ignore}[1]{}
\renewcommand{\theequation}{\arabic{section}.\arabic{equation}}
\newcommand{\Romannumeral}[1]{\uppercase\expandafter{\romannumeral#1}}

\allowdisplaybreaks[4]

\begin{document}

\baselineskip=20pt

\newfont{\elevenmib}{cmmib10 scaled\magstep1}
\newcommand{\preprint}{
     \begin{flushleft}
       \elevenmib Yukawa\, Institute\, Kyoto\\
     \end{flushleft}\vspace{-1.3cm}
     \begin{flushright}\normalsize \sf
       DPSU-09-2\\
       YITP-09-14\\
       March 2009
     \end{flushright}}
\newcommand{\Title}[1]{{\baselineskip=26pt
     \begin{center} \Large \bf #1 \\ \ \\ \end{center}}}
\newcommand{\Author}{\begin{center}
     \large \bf Satoru Odake${}^a$ and Ryu Sasaki${}^b$ \end{center}}
\newcommand{\Address}{\begin{center}
       $^a$ Department of Physics, Shinshu University,\\
       Matsumoto 390-8621, Japan\\
       ${}^b$ Yukawa Institute for Theoretical Physics,\\
       Kyoto University, Kyoto 606-8502, Japan
     \end{center}}
\newcommand{\Accepted}[1]{\begin{center}
     {\large \sf #1}\\ \vspace{1mm}{\small \sf Accepted for Publication}
     \end{center}}

\preprint
\thispagestyle{empty}
\bigskip\bigskip\bigskip

\Title{Unified theory of exactly and quasi-exactly solvable `Discrete'
quantum mechanics:\\ \Romannumeral{1}. Formalism}
\Author

\Address
\vspace{1cm}

\begin{abstract}
We present a simple recipe to construct exactly and quasi-exactly
solvable Hamiltonians in one-dimensional `discrete' quantum mechanics,
in which the Schr\"{o}dinger equation is a difference equation. 
It reproduces all the known ones whose eigenfunctions consist of
the Askey scheme of hypergeometric orthogonal polynomials of a
continuous or a discrete variable.
The recipe also predicts several new ones. An essential role is
played by the sinusoidal coordinate, which generates the closure
relation and the Askey-Wilson algebra together with the Hamiltonian.
The relationship between the closure relation and the Askey-Wilson
algebra is clarified.
\end{abstract}

\section{Introduction}
\label{sec:intro}

For one dimensional quantum mechanical systems, two sufficient
conditions for {\em exact solvability\/} are known.
The first is the {\em shape invariance\/} \cite{genden}, which
guarantees exact solvability in the Sch\"odinger picture.  
The whole set of energy eigenvalues and the corresponding
eigenfunctions can be obtained explicitly through shape invariance
combined with Crum's theorem \cite{crum}, or the factorisation
method \cite{infhul} or the supersymmetric quantum mechanics \cite{susyqm}.
The second is the {\em closure relation\/} \cite{os7}. It allows
to construct the exact Heisenberg operator solution of the
{\em sinusoidal coordinate\/} $\eta(x)$, which generates the closure
relation together with the Hamiltonian.
The positive/negative energy parts of the Heisenberg operator
solution give the {\em annihilation}/{\em creation operators\/},
in terms of which every eigenstate can be built up algebraically
starting from the groundstate. Thus exact solvability in the Heisenberg
picture is realised.

It is interesting to note that these two sufficient conditions apply
equally well in the `discrete' quantum mechanics (QM)
\cite{os4,os5,os7,os12,os13}, which is a simple extension or
deformation of QM.
In discrete QM the dynamical variables are, as in the ordinary QM,
the coordinate $x$ and the conjugate momentum $p$, which is realised
as $p=-i\partial_x$.
The Hamiltonian contains the momentum operator in exponentiated
forms $e^{\pm\beta p}$, which acts on wavefunctions as
{\em finite shift operators\/}, either in the pure imaginary
directions or the real directions.
Thus the Schr\"odinger equation in discrete QM is a
{\em difference equation\/} instead of differential in ordinary QM.
Various examples of exactly solvable discrete quantum mechanics are
known for both of the two types of shifts
\cite{os4,os5,os6,os7,os12,os13}, and the eigenfunctions
consist of the Askey-scheme of the hypergeometric orthogonal
polynomials \cite{askey,ismail,koeswart} of a continuous
(pure imaginary shifts) and a discrete (real shifts) variable.

It should be stressed, however, that these two sufficient conditions
do not tell how to build exactly solvable models.
In this paper we present a simple theory of constructing exactly
solvable Hamiltonians in discrete QM.
It covers all the known examples of exactly solvable discrete QM
with both pure imaginary and real shifts \cite{os12,os13} and
it predicts several new ones to be explored in a subsequent
publication \cite{os16}.
Moreover, the theory is general enough to generate
{\em quasi-exactly solvable\/} Hamiltonians in the same manner.
The quasi-exact solvability means, in contrast to the exact solvability,
that only a finite number of energy eigenvalues and the corresponding
eigenfunctions can be obtained exactly \cite{Ush}. 
This unified theory also incorporates the known examples of
quasi-exactly solvable Hamiltonians \cite{os10,newqes}. 
A new type of quasi-exactly solvable Hamiltonians is constructed
in this paper and its explicit examples will be surveyed in a
subsequent publication \cite{os16}.
One of the merits of the present approach is that it reveals
the common structure underlying the exactly and quasi-exactly
solvable theories.
In ordinary QM, the corresponding theory was already given
in the Appendix A of \cite{os7}, although it does not cover the
quasi-exact solvability.

The present paper is organised as follows.
In section two the general setting of the discrete quantum mechanics
is briefly reviewed and in \S\ref{sec:H} the Hamiltonians for
the pure imaginary shifts and for the real shifts cases are given
and the general strategy of working in the vector space of polynomials
in the sinusoidal coordinate is explained.
In \S\ref{sec:eta}, based on a few postulates, various properties
of the sinusoidal coordinate $\eta(x)$, which is the essential
ingredient of the present theory, are presented in some detail.
The main result of the paper, the unified form of the exactly and
quasi-exactly solvable `Hamiltonians,\!\!' is given in \S\ref{sec:pot_fn}.
The action of the Hamiltonian on the polynomials of the sinusoidal
coordinate is explained in \S\ref{sec:Ht_on_etan}.
It simply maps a degree $n$ polynomial into a degree $n+L-2$ polynomial.
Here $L$ is the degree of a certain polynomial constituting the
potential function in the Hamiltonian.
The exactly solvable case ($L=2$) is discussed in section three.
In \S\ref{sec:closurerelation}, the closure relation, which used
to be verified for each given Hamiltonian, is shown to be satisfied
once and for all by the proposed exactly solvable Hamiltonian.
The nature of the dual closure relation, which plays an important
role in the theory of discrete QM with real shifts and the
corresponding theory of orthogonal polynomials of a discrete variable,
is examined and compared with that of the closure relation
in \S\ref {sec:dualclosurerelation}.
The relationship between the closure plus dual closure relations
and the Askey-Wilson algebra \cite{zhedanov,GLZ,vinzhed,terw} is
 elucidated in \S\ref{AW3}.
In \S\ref{sec:shapeinvariance}, shape invariance is explained and
shown to be satisfied for the pure imaginary shifts case
\S\ref{sec:shapeinv_cont} and for the real shifts case
\S\ref{sec:shapeinv_dics}. The quasi-exactly solvable `Hamiltonians'
are discussed in section four. The QES case with $L=3$ is achieved
in \S\ref{sec:L=3} by adjusting the compensation term which is
linear in $\eta(x)$. A new type of QES with $L=4$ is introduced
in \S\ref{sec:L=4}, which has quadratic in $\eta(x)$ compensation terms.
It is shown that QES is not possible for $L\ge 5$ in \S\ref{sec:L>=5}.
The issue of returning from the `Hamiltonian' in the polynomial
space to the original Hamiltonian $\mathcal{H}$ is discussed
in section five. This is related to the properties of the
(pseudo-)groundstate $\phi_0$. The final section is for a summary,
containing the simple recipe to construct exactly and quasi-exactly
solvable Hamiltonians.
Appendix \ref{app:sinusoidal} provides the explicit forms of the
sinusoidal coordinates with which the actual exactly and quasi-exactly
solvable Hamiltonians are constructed.
There are eight different $\eta(x)$ for the continuous variable $x$
and five for the discrete $x$.
Appendix \ref{app:hermiticity} gives the proof of the hermiticity of
the Hamiltonian, which is slightly more involved than in the ordinary QM.
Appendix \ref{app:upper} recapitulates the elementary formulas for
the eigenvalues and eigenvectors of an upper-triangular matrix,
to which the exactly solvable ($L=2$) `Hamiltonian' in the polynomial
space reduces.
 
\section{`Discrete' Quantum Mechanics}
\label{discQM}
\setcounter{equation}{0}

Throughout this paper we consider `discrete' quantum mechanics of
one degree of freedom. Discrete quantum mechanics is a generalisation
of quantum mechanics in which the Sch\"{o}dinger equation is a
difference equation instead of differential in ordinary QM
\cite{os4,os5,os6,os7,os12,os13}.
In other words, the Hamiltonian contains the momentum operator
$p=-i\partial_x$ in exponentiated forms $e^{\pm\beta p}$ which
work as shift operators on the wavefunction
\begin{equation}
  e^{\pm\beta p}\psi(x)=\psi(x\mp i\beta).
\end{equation}
According to the two choices of the parameter $\beta$,
either {\em real\/} or {\em pure imaginary\/}, we have two types
of discrete QM;
with (\romannumeral1) pure imaginary shifts, or (\romannumeral2)
real shifts, respectively.
In the case of pure imaginary shifts,
$\psi(x\mp i\gamma)$, $\gamma\in\mathbb{R}_{\neq0}$, we require
the wavefunction to be an {\em analytic\/} function of $x$ with
its domain including the real axis or a part of it on which the
dynamical variable $x$ is defined. For the real shifts case,
the difference equation gives constraints on wavefunctions only
on equally spaced lattice points. Then we choose, after proper
rescaling, the variable $x$ to be an {\em integer\/}, with the
total number either finite ($N+1$) or infinite.

To sum up, the dynamical variable $x$ of the one dimensional
discrete quantum mechanics takes continuous or discrete values:
\begin{align}
  \textit{imaginary shifts}&:\quad x\in\mathbb{R},\quad x\in(x_1,x_2),
  \label{xconti}\\
  \textit{real shifts}&:\quad x\in\mathbb{Z},\quad x\in[0,N]\text{  or  }
  [0,\infty).
  \label{xdiscr}
\end{align}
Here $x_1$, $x_2$ may be finite, $-\infty$ or  $+\infty$.
Correspondingly, the inner product of the wavefunctions has the
following form:
\begin{align}
  \textit{imaginary shifts}&:\quad (f,g)=\int_{x_1}^{x_2}f^*(x)g(x)dx,
  \label{inn_pro}\\
  \textit{real shifts}&:\quad (f,g)=\sum_{x=0}^Nf(x)^*g(x)
  \ \text{ or }\ \sum_{x=0}^{\infty}f(x)^*g(x),
\end{align}
and the norm of $f(x)$ is $|\!|f|\!|=\sqrt{(f,f)}$.
In the case of imaginary shifts, other functions appearing
in the Hamiltonian need to be {\em analytic\/} in $x$ within
the same domain.
Let us introduce the $*$-operation on an analytic function,
$*:f\mapsto f^*$.
If $f(x)=\sum\limits_{n}a_nx^n$, $a_n\in\mathbb{C}$, then
$f^*(x)\eqdef\sum\limits_{n}a_n^*x^n$, in which $a_n^*$ is the complex
conjugation of $a_n$. Obviously $f^{**}(x)=f(x)$ and $f(x)^*=f^*(x^*)$.
\label{star}
If $f$ is an analytic function, so is $g(x)\eqdef f(x-a)$,
$a\in\mathbb{C}$. The $*$-operation on this analytic function is
$g^*(x)=\bigl(f(x^*-a)\bigr)^*=f^*(x-a^*)$.
If a function satisfies $f^*=f$, then it takes real values on the
real line.
The `absolute value' of an analytic function to be used in this paper
is defined by $|f(x)|\eqdef\sqrt{f(x)f^*(x)}$, which is again analytic
and real non-negative on the real axis.
Note that the $*$-operation is used in the inner product for
the pure imaginary shifts case \eqref{inn_pro} so that the entire
integrand is an analytic function, too.
This is essential for the proof of hermiticity to be presented
in Appendix \ref{app:hermiticity}.

In quantum mechanics, the eigenvalue problem of a given Hamiltonian
is the central issue. In this paper, we will consider the Hamiltonians
having finite or semi-infinite discrete energy levels only:
\begin{equation}
  0=\mathcal{E}(0)<\mathcal{E}(1)<\mathcal{E}(2)<\cdots.
  \label{positivesemi}
\end{equation}
Here we have chosen the additive constant of the Hamiltonian so that
the groundstate energy vanishes. In other words, the Hamiltonian is
{\em positive semi-definite}. It is a well known theorem in linear
algebra that any positive semi-definite hermitian matrix can be
factorised as a product of a certain matrix, say $\mathcal{A}$, and
its hermitian conjugate $\mathcal{A^\dagger}$.
As we will see shortly, the Hamiltonians of discrete quantum mechanics
have the same property, both with the imaginary and real shifts.

\subsection{Hamiltonian and Strategy}
\label{sec:H}

The Hamiltonian of one dimensional discrete quantum mechanics has
a simple form 
\begin{equation}
  \mathcal{H}\eqdef\varepsilon\Bigl(
  \sqrt{V_+(x)}\,e^{\beta p}\sqrt{V_-(x)}
  +\!\sqrt{V_-(x)}\,e^{-\beta p}\sqrt{V_+(x)}
  -V_+(x)-V_-(x)\Bigr).
  \label{H}
\end{equation}
Corresponding to the imaginary/real shifts cases, the parameter $\beta$,
the potential functions $V_{\pm}(x)$ and a sign factor $\varepsilon$ are
\begin{alignat}{5}
  \textit{imaginary shifts}&:\quad& \beta&=\gamma,\quad&
  \varepsilon&=1,\quad& V_+(x)&=V(x), \quad& V_-(x)&=V^*(x),\n
  \textit{real shifts}&:\quad& \beta&=i,\quad&
  \varepsilon&=-1,\quad& V_+(x)&=B(x), \quad& V_-(x)&=D(x),
\end{alignat}
with $\gamma\in\mathbb{R}_{\neq0}$.
The potential function $B(x)$ and $D(x)$ are positive and vanish
at boundaries:
\begin{equation}
  B(x)>0,\quad D(x)>0,\quad  D(0)=0\ ;\quad
  B(N)=0\ \ \text{for the finite case}.
  \label{D(0)=0}
\end{equation}
As mentioned above, $e^{\pm\beta p}$ are shift operators
$e^{\pm\beta p}f(x)=f(x\mp i\beta)$, and the Schr\"odinger equation
\begin{equation}
  \mathcal{H}\phi_n(x)=\mathcal{E}(n)\phi_n(x),\qquad
  n=0,1,2,\ldots,
  \label{Sch_eq}
\end{equation}
is a difference equation.
The hermiticity of the Hamiltonian is manifest for the real shifts
case because the Hamiltonian is a real symmetric matrix.
For the imaginary shifts case, see Appendix \ref{app:hermiticity}.

This positive semi-definite Hamiltonian \eqref{H} can be factorized:
\begin{equation}
  \mathcal{H}=\mathcal{A}^{\dagger}\mathcal{A}.
  \label{factor}
\end{equation}
Corresponding to the imaginary/real shifts cases, $\mathcal{A}$ and
$\mathcal{A}^{\dagger}$ are
\begin{gather}
  \mathcal{A}=i\bigl(e^{\gamma p/2}\sqrt{V^*(x)}
  -e^{-\gamma p/2}\sqrt{V(x)}\bigr),\quad
  \mathcal{A}^{\dagger}=-i\bigl(\sqrt{V(x)}\,e^{\gamma p/2}
  -\sqrt{V^*(x)}\,e^{-\gamma p/2}\bigr),\\
  \mathcal{A}=\sqrt{B(x)}-e^{\partial}\sqrt{D(x)},\quad
  \mathcal{A}^{\dagger}=\sqrt{B(x)}-\sqrt{D(x)}\,e^{-\partial}.
\end{gather}
The groundstate wavefunction $\phi_0(x)$ is determined as a zero mode
of $\mathcal{A}$,
\begin{equation}
  \mathcal{A}\phi_0(x)=0.
  \label{Aphi0=0}
\end{equation}
The similarity transformed Hamiltonian $\widetilde{\mathcal{H}}$
in terms of the groundstate wavefunction $\phi_0$ has a much simpler
form than the original Hamiltonian $\mathcal{H}$:
\begin{align}
  \widetilde{\mathcal{H}}&\eqdef
  \phi_0(x)^{-1}\circ \mathcal{H}\circ\phi_0(x)
  \label{Htdef}\\
  &=\varepsilon\Bigl(V_+(x)(e^{\beta p}-1)+V_-(x)(e^{-\beta p}-1)\Bigr).
  \label{Ht}
\end{align}
In the second equation we have used \eqref{Aphi0=0}.

In the following we will take $\widetilde{\mathcal{H}}$ instead of
$\mathcal{H}$ as the starting point. That is, we reverse the argument
and construct directly the `Hamiltonian' $\widetilde{\mathcal{H}}$
\eqref{Ht} based on a certain function $\eta(x)$ to be called
the {\em sinusoidal coordinate\/}. The necessary properties of
the sinusoidal coordinate will be introduced in the next subsection
\S\ref{sec:eta}.
The general strategy is to construct the `Hamiltonian'
$\widetilde{\mathcal{H}}$ in such a way that
it maps a polynomial in $\eta(x)$ into another:
$\widetilde{\mathcal{H}}\mathcal{V}_n \subseteq
\mathcal{V}_{n+L-2}\subset\mathcal{V}_{\infty}$.
Here $\mathcal{V}_n$ ($n\in\mathbb{Z}_{\geq 0}$) is defined by
\begin{equation}
  \mathcal{V}_n\eqdef
  \text{Span}\bigl[1,\eta(x),\ldots,\eta(x)^n\bigr], \qquad
  \mathcal{V}_{\infty}\eqdef\lim_{n\to\infty} \mathcal{V}_n.
  \label{Vndef}
\end{equation}
The goal is achieved by choosing very special forms of $V_\pm(x)$
as given in \eqref{V+-=V+-t/etaeta}--\eqref{V+-t}, that is $V(x)$
and $V^*(x)$ or $B(x)$ and $D(x)$ are polynomials of degree $L$ in
the sinusoidal coordinate $\eta(x)$ and its shifts $\eta(x\mp i\beta)$
divided by special quadratic polynomials in them.
This provides a unified theory of exactly solvable and quasi-exactly
solvable discrete QM.
Exactly solvable QM are realised by choosing $\widetilde{\mathcal{H}}$
in such a way ($L=2$) that
$\widetilde{\mathcal{H}}\mathcal{V}_n\subseteq\mathcal{V}_n$ is
satisfied for all $n$.
Then the existence of an eigenfunction, or to be more precise,
a degree $n$ eigenpolynomial, of $\widetilde{\mathcal{H}}$ is
guaranteed for each integer $n$.
On the other hand, quasi-exact solvability is attained by adjusting
the parameters of $\widetilde{\mathcal{H}}$ in such a way ($L=3,4$) that
$\widetilde{\mathcal{H}}'\mathcal{V}_M\subseteq\mathcal{V}_M$ is realised
for an integer $M$. Here $\widetilde{\mathcal{H}}'$ is a modification of
$\widetilde{\mathcal{H}}$ by the addition of the compensation terms.
Then the `Hamiltonian' $\widetilde{\mathcal{H}}'$ has an
$M+1=\mbox{dim}(\mathcal{V}_M)$-dimensional invariant space,
providing $M+1$ eigenpolynomials of $\widetilde{\mathcal{H}}'$.
After obtaining such (quasi-)exactly solvable `Hamiltonian'
$\widetilde{\mathcal{H}}$ ($\widetilde{\mathcal{H}}'$), we have to
find the (quasi-)groundstate wavefunction $\phi_0$ in order to return
to the true Hamiltonian $\mathcal{H}$ ($\mathcal{H}'$) by \eqref{Htdef}.
It should be noted that the existence of such a (quasi-)groundstate
wavefunction is not guaranteed {\it a priori\/} since we have started
with $\widetilde{\mathcal{H}}$ instead of $\mathcal{H}$.
In the case of quasi-exactly solvable QM, the positive semi-definiteness
of the Hamiltonian \eqref{positivesemi} is in general lost due to
the inclusion of the compensation terms to $\widetilde{\mathcal{H}}'$.

\subsection{sinusoidal coordinate}
\label{sec:eta}

Motivated by the study in \cite{os7,os12,os13}, let us define a
sinusoidal coordinate $\eta(x)$ as a real (or `real' analytic
$\eta^*(x)=\eta(x)$ in the case of pure imaginary shifts) function
of $x$ satisfying the following symmetric shift-addition property:
\begin{equation}
  \eta(x-i\beta)+\eta(x+i\beta)=(2+r_1^{(1)})\eta(x)+r_{-1}^{(2)}.
  \label{eta1+etam1}
\end{equation}
Here $r_1^{(1)}$ and $r_{-1}^{(2)}$ are real parameters and we assume
$r_1^{(1)}>-4$.
These two, $r_1^{(1)}$ and $r_{-1}^{(2)}$, are fundamental parameters
appearing in both exactly and quasi-exactly solvable dynamical systems.
For the exactly solvable systems, these two parameters also manifest
themselves \eqref{Ricoeff} in the {\em closure relation\/}, another
characterisation of exact solvability, to be discussed in
\S\ref{sec:closurerelation}.
Since a polynomial in $\eta(x)$ is also a polynomial in $a\eta(x)+b$
($a,b$: real constants), we impose two conditions\footnote{
For the real shifts case, such $\eta(x)$ satisfying \eqref{eta1+etam1}
and \eqref{eta(0)=0} can be classified into five types
\eqref{etaform1'}--\eqref{etaform5'} \cite{os12}
and they also satisfy the condition \eqref{eta1*etam1}.
}
(we assume $0\in[x_1,x_2]$)
\begin{equation}
  \eta(0)=0\quad \text{and}\quad
  \eta(x)\text{ : monotone increasing function},
  \label{eta(0)=0}
\end{equation}
which are not essential for (quasi-)exact solvability but important for
expressing various formulas in a unified way.
We impose another condition, to be called the symmetric
shift-multiplication property:
\begin{equation}
  \eta(x-i\beta)\eta(x+i\beta)
  =\bigl(\eta(x)-\eta(-i\beta)\bigr)\bigl(\eta(x)-\eta(i\beta)\bigr),
  \label{eta1*etam1}
\end{equation}
together with $\eta(x)\neq\eta(x-i\beta)\neq\eta(x+i\beta)\neq\eta(x)$.

The two conditions \eqref{eta1+etam1} and \eqref{eta1*etam1} imply
that any symmetric polynomial in $\eta(x-i\beta)$ and $\eta(x+i\beta)$
is expressed as a polynomial in $\eta(x)$. Especially we have ($n\geq -1$)
\begin{equation}
  g_n(x)\eqdef\frac{\eta(x-i\beta)^{n+1}-\eta(x+i\beta)^{n+1}}
  {\eta(x-i\beta)-\eta(x+i\beta)}
  =\genfrac{(}{)}{0pt}{}
  {\text{a polynomial of}}{\text{degree $n$ in $\eta(x)$}}
  =\sum_{k=0}^ng_n^{(k)}\eta(x)^{n-k}.
\end{equation}
The coefficient $g_n^{(k)}$ is real because $g^*_n(x)=g_n(x)$.
We set $g_n^{(k)}=0$ except for $0\leq k\leq n$.
Since $g_n(x)$ satisfies the following three term recurrence relation
\begin{equation}
  g_{n+1}(x)=\bigl(\eta(x-i\beta)+\eta(x+i\beta)\bigr)g_n(x)
  -\eta(x-i\beta)\eta(x+i\beta)g_{n-1}(x)\quad(n\geq 0),
\end{equation}
we can write down $g_n^{(k)}$ explicitly.
Especially $g_n^{(k)}$ for $k=0,1$ are
\begin{align}
  g_n^{(0)}&=[n+1],\qquad
  \label{g0n}\\
  g_n^{(1)}&=
  \begin{cases}
   {\displaystyle \tfrac16n(n+1)(2n+1)r_{-1}^{(2)}}
   &\text{for }r_1^{(1)}=0,\\
   {\displaystyle \frac{n[n+1]-(n+1)[n]}{r_1^{(1)}}\,r_{-1}^{(2)}}
   &\text{for }r_1^{(1)}\neq 0.
  \end{cases}
  \label{g1n}
\end{align}
Here we have defined $[n]$ as
\begin{equation}
  [n]\eqdef
  \begin{cases}
   n&\text{for }r_1^{(1)}=0,\\
   {\displaystyle
   \frac{e^{\alpha n}-e^{-\alpha n}}{e^{\alpha}-e^{-\alpha}}}
   &\text{for }r_1^{(1)}>0\ \ \bigl(\Leftarrow
   r_1^{(1)}=(e^{\frac{\alpha}{2}}-e^{-\frac{\alpha}{2}})^2
   \ \ (\alpha>0)\bigr),\\
   {\displaystyle
   \frac{e^{i\alpha n}-e^{-i\alpha n}}{e^{i\alpha}-e^{-i\alpha}}}
   &\text{for }-4<r_1^{(1)}<0\ \ \bigl(\Leftarrow
   r_1^{(1)}=(e^{i\frac{\alpha}{2}}-e^{-i\frac{\alpha}{2}})^2
   \ \ (0<\alpha<\pi)\bigr).
  \end{cases}
  \label{[n]}
\end{equation}
Note that $r_1^{(1)}$ and $r_{-1}^{(2)}$ are expressed as
\begin{equation}
  r_1^{(1)}=[2]-2,\quad r_{-1}^{(2)}=\eta(-i\beta)+\eta(i\beta).
  \label{r11_rm12}
\end{equation}
For $n,m\in\mathbb{Z}$, $n\geq m-1$, we have
\begin{equation}
   \sum_{r=m}^ng_r^{(1)}=
  \begin{cases}
   {\displaystyle \tfrac{1}{12}(n+m+1)(n-m+1)(n^2+2n+m^2)\,r_{-1}^{(2)}}
   &\text{for } r_1^{(1)}=0,\\[6pt]
   {\displaystyle \frac{(n+1)[n+1]-m[m]-[\frac12]^{-2}[\frac{n+m+1}{2}]
   [\frac{n-m+1}{2}]}{r_1^{(1)}}\,r_{-1}^{(2)}}
   &\text{for } r_1^{(1)}\neq 0.
  \end{cases}
  \label{sumg1}
\end{equation}
The following properties of $[n]$ are useful:
\begin{align}
  &[a][a+c]-[b][b+c]=[a-b][a+b+c],
  \label{[a][a+c]..}\\[4pt]
  &\sum_{r=m}^n\,[r]=\frac{[\frac{n+m}{2}][\frac{n-m+1}{2}]}{[\frac12]}
  \quad(n,m\in\mathbb{Z},\ n\geq m-1).
  \label{sum[r]}
\end{align}

\subsection{potential functions}
\label{sec:pot_fn}

The first goal is to construct a general form of the `Hamiltonian'
$\widetilde{\mathcal{H}}$ such that a polynomial in $\eta(x)$ is
mapped into another.
It is achieved by the following form of the potential functions
$V_{\pm}(x)$:
\begin{align}
  V_{\pm}(x)&=\frac{\widetilde{V}_{\pm}(x)}
  {\bigl(\eta(x\mp i\beta)-\eta(x)\bigr)
  \bigl(\eta(x\mp i\beta)-\eta(x\pm i\beta)\bigr)}\,,
  \label{V+-=V+-t/etaeta}\\[4pt]
  \widetilde{V}_{\pm}(x)&=\sum_{\genfrac{}{}{0pt}{}{k,l\geq 0}{k+l\leq L}}
  v_{k,l}\,\eta(x)^k\eta(x\mp i\beta)^l,
  \label{V+-t}
\end{align}
where $L$ is a natural number roughly indicating the degree of $\eta(x)$ in 
$\widetilde{V}_{\pm}(x)$ and $v_{k,l}$ are real constants,
with the constraint $\sum\limits_{k+l=L}v_{k,l}^2\neq0$.
It is important that the same $v_{k,l}$ appears in both
$\widetilde{V}_{\pm}(x)$.
As we will see in the next subsection \S\ref{sec:Ht_on_etan},
the `Hamiltonian' $\widetilde{\mathcal{H}}$ with the above $V_\pm(x)$
maps a degree $n$ polynomial in $\eta(x)$ to a degree $n+L-2$
polynomial \eqref{Htetamn}, \eqref{hvnhvnl}.

The essential part of the formula \eqref{V+-=V+-t/etaeta} is the
denominators. They have the same form as the generic formula,
derived by the present authors, for the coefficients of the three term
recurrence relations of the orthogonal polynomials, (4.52) and (4.53)
in \cite{os12}.
The translation rules are the {\em duality\/} correspondence itself,
(3.14)--(3.18) in \cite{os12}:
\begin{gather}
  \mathcal{E}(n)\to\eta(x),\qquad -A_n\to V_+(x),
  \qquad -C_n\to V_-(x),\n
  \alpha_+\bigl(\mathcal{E}(n)\bigr)\to \eta(x-i\beta)-\eta(x),\quad
  \alpha_-\bigl(\mathcal{E}(n)\bigr)\to \eta(x+i\beta)-\eta(x).
\end{gather}

Some of the parameters $v_{k,l}$ in \eqref{V+-t} are redundant.
{}From \eqref{eta1+etam1} and \eqref{eta1*etam1}, we have
\begin{equation}
  \eta(x\mp i\beta)^2=(2+r_1^{(1)})\eta(x)\eta(x\mp i\beta)-\eta(x)^2
  +r_{-1}^{(2)}\bigl(\eta(x)+\eta(x\mp i\beta)\bigr)-\eta(-i\beta)\eta(i\beta).
\end{equation}
By using this repeatedly, a monomial $\eta(x\mp i\beta)^l$ can be
reduced to a polynomial of degree one in $\eta(x\mp i\beta)$ whose
coefficients are polynomials in $\eta(x)$.
Therefore it is sufficient to keep $v_{k,l}$ with $l=0,1$.
The remaining $2L+1$ parameters $v_{k,l}$ ($k+l\leq L$, $l=0,1$) are
independent, with one of which corresponds to the overall normalization
of the Hamiltonian.
In fact, if two sets of parameters $\{v_{k,l}\}$ and $\{v'_{k,l}\}$
($k+l\leq L$, $l=0,1$) give the same $V_{\pm}(x)$, namely,
$\sum_{k=0}^L(v_{k,0}-v'_{k,0})\eta(x)^k
+\sum_{k=0}^{L-1}(v_{k,1}-v'_{k,1})\eta(x)^k\eta(x\mp i\beta)=0$,
then we obtain $v_{k,l}=v'_{k,l}$. Therefore there is no more
redundancy in $v_{k,l}$ ($k+l\leq L$, $l=0,1$).
Note that we have not yet imposed the boundary condition $D(0)=0$
\eqref{D(0)=0}.
The sinusoidal coordinate $\eta(x)$ itself may have extra parameters.

\subsection{$\widetilde{\mathcal{H}}$ on the polynomial space}
\label{sec:Ht_on_etan}

The action of $\widetilde{\mathcal{H}}$ \eqref{Ht} on $\eta(x)^n$ becomes
with \eqref{V+-=V+-t/etaeta} and \eqref{V+-t}:
\begin{align}
  \widetilde{\mathcal{H}}\eta(x)^n
  &=\varepsilon\Bigl(
  V_+(x)\bigl(\eta(x-i\beta)^n-\eta(x)^n\bigr)
  +V_-(x)\bigl(\eta(x+i\beta)^n-\eta(x)^n\bigr)\Bigr)\n
  &=\varepsilon\,
  \frac{\widetilde{V}_+(x)\sum_{r=0}^{n-1}\eta(x)^r\eta(x-i\beta)^{n-1-r}
  -\widetilde{V}_-(x)\sum_{r=0}^{n-1}\eta(x)^r\eta(x+i\beta)^{n-1-r}}
  {\eta(x-i\beta)-\eta(x+i\beta)}\n
  &=\varepsilon \sum_{r=0}^{n-1}\eta(x)^r\,
  \frac{\widetilde{V}_+(x)\eta(x-i\beta)^{n-1-r}
  -\widetilde{V}_-(x)\eta(x+i\beta)^{n-1-r}}
  {\eta(x-i\beta)-\eta(x+i\beta)}\n
  &=\varepsilon \sum_{r=0}^{n-1}\eta(x)^r
  \sum_{\genfrac{}{}{0pt}{}{k,l\geq 0}{k+l\leq L}}v_{k,l}\,\eta(x)^k\,
  \frac{\eta(x-i\beta)^{l+n-1-r}-\eta(x+i\beta)^{l+n-1-r}}
  {\eta(x-i\beta)-\eta(x+i\beta)}\n
  &=\varepsilon \sum_{\genfrac{}{}{0pt}{}{k,l\geq 0}{k+l\leq L}}v_{k,l}
  \sum_{r=0}^{n-1}\eta(x)^{k+r}g_{n+l-r-2}(x)\n
  &=\bigl(\text{a polynomial of degree $n+L-2$ in $\eta(x)$}\bigr)\n
  &=\varepsilon \sum_{\genfrac{}{}{0pt}{}{k,l\geq 0}{k+l\leq L}}
  \sum_{r=0}^{n-1}\sum_{j=0}^{n+l-r-2}v_{k,l}\,g_{n+l-r-2}^{(j)}\,
  \eta(x)^{n+k+l-2-j}\n
  &=\varepsilon
  \sum_{m=0}^{n+L-2}\eta(x)^{n+L-2-m}\!\!\!\!\sum_{j=\max(m-L,0)}^m
  \sum_{\genfrac{}{}{0pt}{}{k,l\geq 0}{k+l=L-m+j}}\!\!\!\!v_{k,l}
  \sum_{r=0}^{n-1}g_{n+l-r-2}^{(j)}\n
  &=\sum_{m=0}^{n+L-2}\eta(x)^{n+L-2-m}\!\!\!\!
  \sum_{j=\max(m-L,0)}^m\!\!\!\!e_{m,j,n}.
  \label{Htetan}
\end{align}
Here $e_{m,j,n}$ (the $L$-dependence is implicit) is defined by
\begin{equation}
  e_{m,j,n}\eqdef\varepsilon\sum_{l=0}^{L-m+j}\!\!v_{L-m+j-l,l}
  \sum_{r=0}^{n-1}g_{n+l-r-2}^{(j)}.
  \label{emjn}
\end{equation}
Therefore the matrix elements of $\widetilde{\mathcal{H}}$ in the
basis $\{\eta(x)^n\}_{n=0,1,\ldots}$ is given by
\begin{equation}
  \widetilde{\mathcal{H}}\eta(x)^n
  =\sum_{m=0}^{n+L-2}\eta(x)^m\widetilde{\mathcal{H}}_{m,n}^{\eta},\qquad
  \widetilde{\mathcal{H}}_{m,n}^{\eta}=\!\!\!\!\!
  \sum_{j=\max(n-2-m,0)}^{n+L-2-m}\!\!\!e_{n+L-2-m,j,n}.
  \label{Htetamn}
\end{equation}
The coefficients $e_{m,0,n}$ and $e_{m,1,n}$ become, by using 
\eqref{sum[r]} and \eqref{sumg1}:
\begin{align}
  e_{m,0,n}&=\varepsilon\frac{[\frac{n}{2}]}{[\frac12]}
  \sum_{l=0}^{L-m}v_{L-m-l,l}[\tfrac{n+2l-1}{2}],
  \label{e_{m,0,n}}\\
  e_{m,1,n}&=\varepsilon\sum_{l=0}^{L-m+1}v_{L-m+1-l,l}\n
  &\times
  \begin{cases}
   {\displaystyle \tfrac{1}{12}n(n+2l-2)\bigl((n+l-1)^2+l^2-2l\bigr)
   r_{-1}^{(2)}}
   &\text{for }r_1^{(1)}=0,\\
   {\displaystyle \frac{(n+l-1)[n+l-1]-(l-1)[l-1]-[\frac12]^{-2}
   [\frac{n+2l-2}{2}][\frac{n}{2}]}{r_1^{(1)}}\,r_{-1}^{(2)}}
   &\text{for }r_1^{(1)}\neq 0.
  \end{cases}
  \label{e_{m,1,n}}
\end{align}
So far the conditions $v_{k,l}=0$ for $l\geq 2$ are not used.

We have established
\begin{equation}
  \widetilde{\mathcal{H}}\mathcal{V}_n\subseteq\mathcal{V}_{n+L-2},
  \label{hvnhvnl}
\end{equation}
where $\mathcal{V}_n$ is the polynomial space defined in \eqref{Vndef}.
For $L=2$, $\mathcal{V}_n$ is $\widetilde{\mathcal{H}}$-invariant.
Therefore this case is exactly solvable; all the eigenvalues and
eigenfunctions of $\widetilde{\mathcal{H}}$ can be obtained explicitly
and the eigenfunction is a polynomial of degree $n$ in $\eta(x)$ for
each $n$.
On the other hand, $L\geq 3$ cases are not exactly solvable but some
cases can be made quasi-exactly solvable by certain modification
to be discussed presently.
For $L=0,1$ cases, the matrix $\widetilde{\mathcal{H}}^{\eta}
=(\widetilde{\mathcal{H}}^{\eta}_{m,n})_{0\leq m,n\leq K}$
with finite $K$ is not diagonalizable except for $K=0,1$.

In the following we will set $v_{k,l}=0$ for $l\geq 2$,
see \S\ref{sec:pot_fn}.
For the real shifts case, the condition $D(0)=0$ \eqref{D(0)=0} is
satisfied by choosing $v_{0,0}$ as $v_{0,0}=-v_{0,1}\eta(-1)$.

\section{Exactly Solvable $\widetilde{\mathcal{H}}$}
\label{sec:ES}
\setcounter{equation}{0}

The $L=2$ case is exactly solvable.
Since the Hamiltonian of the polynomial space $\widetilde{\mathcal{H}}$
is an upper triangular matrix \eqref{Htetamn}, its eigenvalues and
eigenvectors are easily obtained explicitly, see Appendix \ref{app:upper}.
The eigenvalue $\mathcal{E}(n)$ is
\begin{equation}
  \mathcal{E}(n)=\widetilde{H}^{\eta}_{n,n}
  =e_{0,0,n}=\varepsilon\frac{[\frac{n}{2}]}{[\frac12]}\Bigl(
  v_{2,0}[\tfrac{n-1}{2}]+v_{1,1}[\tfrac{n+1}{2}]\Bigr),
  \label{energyeigen}
\end{equation}
and the corresponding eigenpolynomial $P_n\bigl(\eta(x)\bigr)$ is expressed
as a determinant of the following order $n+1$ matrix,
\begin{equation}
  P_n\bigl(\eta(x)\bigr)\propto
  \begin{vmatrix}
    1&\eta(x)&\eta(x)^2&\cdots&\eta(x)^n\\
    \mathcal{E}(0)-\mathcal{E}(n)&\widetilde{H}^{\eta}_{0,1}&
    \widetilde{H}^{\eta}_{0,2}&\cdots&\widetilde{H}^{\eta}_{0,n}\\
    &\mathcal{E}(1)-\mathcal{E}(n)&\widetilde{H}^{\eta}_{1,2}&
    \cdots&\widetilde{H}^{\eta}_{1,n}\\
    &&\ddots&\ddots&\vdots\\
    \text{\LARGE $0$}&&&\mathcal{E}(n-1)-\mathcal{E}(n)&
    \widetilde{H}^{\eta}_{n-1,n}
  \end{vmatrix}.
  \label{Pndet}
\end{equation}
For a choice of the sinusoidal coordinate among the possible forms
\eqref{etaform1}--\eqref{etaform5'} and the values of the five parameters,
$v_{0,0}$, $v_{1,0}$, $v_{0,1}$, $v_{1,1}$ and $v_{2,0}$, these two
formulas \eqref{energyeigen} and \eqref{Pndet}, although clumsy, give
the complete solutions of the `Schr\"odinger equation'
$\widetilde{\mathcal{H}}P_n(\eta(x))=\mathcal{E}(n)P_n(\eta(x))$ at
the algebraic level.
For the solutions of a full quantum mechanical problem, however,
one needs the square-integrable groundstate wavefunction $\phi_0(x)$
\eqref{Aphi0=0}, which is essential for the existence of the Hamiltonian
$\mathcal{H}$ and the verification of its hermiticity. These conditions
would usually restrict the ranges of the parameters $v_{0,0},\ldots,v_{2,0}$.

For specific problems, however, there are more powerful and systematic
solution methods based on the shape invariance \cite{genden,os4,os5,os12,os13}
and the closure relation \cite{os7,os12,os13}.
These two are independent and sufficient conditions for exact solvability
which are applicable to not only ordinary QM but also discrete QM.
In our previous works \cite{os4,os5,os7,os12,os13} these conditions were
verified for each specific problem.
Here we will provide proofs based on the generic form of the exactly
solvable ($L=2$) `Hamiltonian' $\widetilde{\mathcal{H}}$, \eqref{Ht},
\eqref{V+-=V+-t/etaeta}, \eqref{V+-t}.
These proofs apply to all the exactly solvable discrete QM.
In the rest of this section we assume the existence of the ground
state wavefunction $\phi_0(x)$ \eqref{Aphi0=0}.

\subsection{closure relation}
\label{sec:closurerelation}

The closure relation is a commutator relation between the Hamiltonian
$\mathcal{H}$ and the sinusoidal coordinate $\eta(x)$ \cite{os7,os12,os13}:
\begin{equation}
  [\mathcal{H},[\mathcal{H},\eta]\,]
  =\eta\,R_0(\mathcal{H})+[\mathcal{H},\eta]\,R_1(\mathcal{H})
  +R_{-1}(\mathcal{H}).
  \label{closurerel}
\end{equation}
Here $R_i(z)$ are polynomials with real coefficients $r_i^{(j)}$,
\begin{equation}
  R_1(z)=r_1^{(1)}z+r_1^{(0)},\quad
  R_0(z)=r_0^{(2)}z^2+r_0^{(1)}z+r_0^{(0)},\quad
  R_{-1}(z)=r_{-1}^{(2)}z^2+r_{-1}^{(1)}z+r_{-1}^{(0)}.
  \label{Ricoeff}
\end{equation}
Reflecting the fact that the Hamiltonian $\mathcal{H}$ has shift operators
$e^{\pm\beta p}$, whereas $\eta(x)$ has none, the function $R_0(\mathcal{H})$
and $R_{-1}(\mathcal{H})$ are quadratic in $\mathcal{H}$ and
$R_{1}(\mathcal{H})$ is linear in $\mathcal{H}$.
By similarity transforming \eqref{closurerel} in terms of the ground
state wavefunction $\phi_0$, it is rewritten as
\begin{equation}
  [\widetilde{\mathcal{H}},[\widetilde{\mathcal{H}},\eta]\,]
  =\eta\,R_0(\widetilde{\mathcal{H}})
  +[\widetilde{\mathcal{H}},\eta]\,R_1(\widetilde{\mathcal{H}})
  +R_{-1}(\widetilde{\mathcal{H}}).
  \label{closurerelt}
\end{equation}
The closure relation \eqref{closurerel} allows us to obtain the exact
Heisenberg operator solution for $\eta(x)$, and the annihilation and
creation operators $a^{(\pm)}$ are extracted from this exact Heisenberg
operator solution \cite{os7}:
\begin{align}
  &e^{it\mathcal{H}}\eta(x)e^{-it\mathcal{H}}
  =a^{(+)}e^{i\alpha_+(\mathcal{H})t}+a^{(-)}e^{i\alpha_-(\mathcal{H})t}
  -R_{-1}(\mathcal{H})R_0(\mathcal{H})^{-1},\\
  &\alpha_{\pm}(\mathcal{H})\eqdef\tfrac12\bigl(R_1(\mathcal{H})
  \pm\sqrt{R_1(\mathcal{H})^2+4R_0(\mathcal{H})}\,\bigr),
  \label{alphapmform}\\
  &\qquad\qquad
  R_1(\mathcal{H})=\alpha_+(\mathcal{H})+\alpha_-(\mathcal{H}),\quad
  R_0(\mathcal{H})=-\alpha_+(\mathcal{H})\alpha_-(\mathcal{H}),\\
  &a^{(\pm)}\eqdef\pm\Bigl([\mathcal{H},\eta(x)]-\bigl(\eta(x)
  +R_{-1}(\mathcal{H})R_0(\mathcal{H})^{-1}\bigr)\alpha_{\mp}(\mathcal{H})
  \Bigr)
  \bigl(\alpha_+(\mathcal{H})-\alpha_-(\mathcal{H})\bigr)^{-1}
  \label{a^{(pm)}}\\
  &\phantom{a^{(\pm)}}=
  \pm\bigl(\alpha_+(\mathcal{H})-\alpha_-(\mathcal{H})\bigr)^{-1}
  \Bigl([\mathcal{H},\eta(x)]+\alpha_{\pm}(\mathcal{H})\bigl(\eta(x)
  +R_{-1}(\mathcal{H})R_0(\mathcal{H})^{-1}\bigr)\Bigr).
\end{align}
The energy spectrum is determined by the over-determined recursion relations
$\mathcal{E}(n+1)=\mathcal{E}(n)+\alpha_+\bigl(\mathcal{E}(n)\bigr)$
and 
$\mathcal{E}(n-1)=\mathcal{E}(n)+\alpha_-\bigl(\mathcal{E}(n)\bigr)$
with $\mathcal{E}(0)=0$, and the excited state wavefunctions $\{\phi_n(x)\}$
are obtained by successive action of the creation operator $a^{(+)}$ on
the groundstate wavefunction $\phi_0(x)$.
The closure relation \eqref{closurerelt} (or \eqref{closurerel}) is
equivalent to the following set of five equations:
\begin{align}
  &\eta(x-2i\beta)-2\eta(x-i\beta)+\eta(x)=r_0^{(2)}\eta(x)+r_{-1}^{(2)}
  +r_1^{(1)}\bigl(\eta(x-i\beta)-\eta(x)\bigr),
  \label{closurerel1}\\
  &\eta(x+2i\beta)-2\eta(x+i\beta)+\eta(x)=r_0^{(2)}\eta(x)+r_{-1}^{(2)}
  +r_1^{(1)}\bigl(\eta(x+i\beta)-\eta(x)\bigr),
  \label{closurerel1p}\\
  &\bigl(\eta(x-i\beta)-\eta(x)\bigr)
  \bigl(V_+(x-i\beta)+V_-(x-i\beta)-V_+(x)-V_-(x)\bigr)\n
  &\quad=-\bigl(r_0^{(2)}\eta(x)+r_{-1}^{(2)}\bigr)
  \bigl(V_+(x-i\beta)+V_-(x-i\beta)+V_+(x)+V_-(x)\bigr)\n
  &\quad\phantom{=}
  -r_1^{(1)}\bigl(\eta(x-i\beta)-\eta(x)\bigr)
  \bigl(V_+(x-i\beta)+V_-(x-i\beta)\bigr)\n
  &\quad\phantom{=}
  +\varepsilon^{-1}\bigl(r_0^{(1)}\eta(x)+r_{-1}^{(1)}
  +r_1^{(0)}\bigl(\eta(x-i\beta)-\eta(x)\bigr)\bigr),
  \label{closurerel2}\\
  &\bigl(\eta(x+i\beta)-\eta(x)\bigr)
  \bigl(V_+(x+i\beta)+V_-(x+i\beta)-V_+(x)-V_-(x)\bigr)\n
  &\quad=-\bigl(r_0^{(2)}\eta(x)+r_{-1}^{(2)}\bigr)
  \bigl(V_+(x+i\beta)+V_-(x+i\beta)+V_+(x)+V_-(x)\bigr)\n
  &\quad\phantom{=}
  -r_1^{(1)}\bigl(\eta(x+i\beta)-\eta(x)\bigr)
  \bigl(V_+(x+i\beta)+V_-(x+i\beta)\bigr)\n
  &\quad\phantom{=}
  +\varepsilon^{-1}\bigl(r_0^{(1)}\eta(x)+r_{-1}^{(1)}
  +r_1^{(0)}\bigl(\eta(x+i\beta)-\eta(x)\bigr)\bigr),
  \label{closurerel2p}\\
  &2\bigl(\eta(x)-\eta(x-i\beta)\bigr)V_+(x)V_-(x-i\beta)
  +2\bigl(\eta(x)-\eta(x+i\beta)\bigr)V_-(x)V_+(x+i\beta)\n
  &\quad=\bigl(r_0^{(2)}\eta(x)+r_{-1}^{(2)}\bigr)
  \bigl(V_+(x)V_-(x-i\beta)+V_-(x)V_+(x+i\beta)+\bigl(V_+(x)+V_-(x)\bigr)^2
  \bigr)\n
  &\quad\phantom{=}
  +r_1^{(1)}\bigl(\eta(x-i\beta)-\eta(x)\bigr)V_+(x)V_-(x-i\beta)
  +r_1^{(1)}\bigl(\eta(x+i\beta)-\eta(x)\bigr)V_-(x)V_+(x+i\beta)\n
  &\quad\phantom{=}
  -\varepsilon^{-1}\bigl(r_0^{(1)}\eta(x)+r_{-1}^{(1)}\bigr)
  \bigl(V_+(x)+V_-(x)\bigr)
  +\varepsilon^{-2}\bigl(r_0^{(0)}\eta(x)+r_{-1}^{(0)}\bigr).
  \label{closurerel3}
\end{align}

Obviously \eqref{closurerel1} and \eqref{closurerel1p} are equivalent
and so are \eqref{closurerel2} and \eqref{closurerel2p},
under the condition \eqref{r02=r11}.
By substituting our choice of $V_{\pm}(x)$
\eqref{V+-=V+-t/etaeta}--\eqref{V+-t} for $L=2$,
it is straightforward to verify the other three equations
\eqref{closurerel2}--\eqref{closurerel3}.
The coefficients $r_i^{(j)}$ appearing in \eqref{Ricoeff} are
expressed by the parameters $v_{1,0}$, $v_{0,1}$, $v_{1,1}$ and
$v_{2,0}$ together with the two parameters $r_1^{(1)}$ and
$r_{-1}^{(2)}$ which have already appeared in the definition of
$\eta(x)$ \eqref{eta1+etam1} (see also \eqref{r11_rm12}):
\begin{alignat}{2}
  r_0^{(2)}&=r_1^{(1)},&\qquad r_0^{(1)}&=2r_1^{(0)},
  \label{r02=r11}\\
  \varepsilon^{-1}r_1^{(0)}&=v_{2,0}+v_{1,1},&\qquad
  \varepsilon^{-2}r_0^{(0)}&=-v_{2,0}\,v_{1,1},
  \label{r10}\\
  \varepsilon^{-1}r_{-1}^{(1)}&=v_{1,0}+v_{0,1},&\qquad
  \varepsilon^{-2}r_{-1}^{(0)}&=-v_{2,0}\,v_{0,1}.
  \label{rm11}
\end{alignat}
Note that $v_{0,0}$ does not appear.
It implies that for the imaginary shifts case the commutation relation
between the annihilation and creation operators does not depend on
$v_{0,0}$. With these formulas, the explicit forms of
$\alpha_\pm(\mathcal{H})$ \eqref{alphapmform} can be expressed in terms
of $r^{(1)}_1$, $v_{2,0}$ and $v_{1,1}$.
It is straightforward to verify the eigenvalue formula \eqref{energyeigen}.
This concludes the unified proof of the closure relation for all the
discrete QM.

\subsection{dual closure relation}
\label{sec:dualclosurerelation}

The {\em dual closure relation\/} has the same forms as the closure
relation \eqref{closurerel} and \eqref{closurerelt} with the roles of
Hamiltonian $\mathcal{H}$ ($\widetilde{\mathcal{H}}$) and the
sinusoidal coordinate $\eta(x)$ exchanged:
\begin{align}
  [\eta,[\eta,\mathcal{H}]\,]&=\mathcal{H}\,R_0^{\text{dual}}(\eta)
  +[\eta,\mathcal{H}]\,R_1^{\text{dual}}(\eta)+R_{-1}^{\text{dual}}(\eta),
  \label{dualclosurerel}\\[4pt]
  [\eta,[\eta,\widetilde{\mathcal{H}}]\,]&=
  \widetilde{\mathcal{H}}\,R_0^{\text{dual}}(\eta)
  +[\eta,\widetilde{\mathcal{H}}]\,R_1^{\text{dual}}(\eta)
  +R_{-1}^{\text{dual}}(\eta),
  \label{dualclosurerelt}
\end{align}
where $R_i^{\text{dual}}(z)$ are as yet unknown polynomials.
We will show below that the dual closure relation is the characteristic
feature shared by all the `Hamiltonians' $\widetilde{\mathcal{H}}$ which
map a polynomial in $\eta(x)$ into another.
Therefore its dynamical contents are not so constraining as the closure
relation, {\em except for\/} the real shifts (the discrete variable)
exactly solvable ($L=2$) case, where the closure relation and the dual
closure relations are on the same footing as shown in \cite{os12}.
By substituting the `Hamiltonian' $\widetilde{\mathcal{H}}$ \eqref{Ht}
without any further specification of $V_\pm$ into the above
\eqref{dualclosurerelt}, we find it is equivalent to the following set
of three equations:
\begin{align}
  &\bigl(\eta(x)-\eta(x-i\beta)\bigr)^2
  =R_0^{\text{dual}}\bigl(\eta(x-i\beta)\bigr)
  +\bigl(\eta(x)-\eta(x-i\beta)\bigr)
  R_1^{\text{dual}}\bigl(\eta(x-i\beta)\bigr),
  \label{dcrcond1}\\
  &\bigl(\eta(x)-\eta(x+i\beta)\bigr)^2
  =R_0^{\text{dual}}\bigl(\eta(x+i\beta)\bigr)
  +\bigl(\eta(x)-\eta(x+i\beta)\bigr)
  R_1^{\text{dual}}\bigl(\eta(x+i\beta)\bigr),
  \label{dcrcond1p}\\
  &0=-\varepsilon\bigl(V_+(x)+V_-(x)\bigr)
  R_0^{\text{dual}}\bigl(\eta(x)\bigr)
  +R_{-1}^{\text{dual}}\bigl(\eta(x)\bigr).
  \label{dcrcond2}
\end{align}
These imply
\begin{align}
  R_1^{\text{dual}}\bigl(\eta(x)\bigr)&=
  \bigl(\eta(x-i\beta)-\eta(x)\bigr)+\bigl(\eta(x+i\beta)-\eta(x)\bigr),
  \label{dualclcon1}\\
  R_0^{\text{dual}}\bigl(\eta(x)\bigr)&=
  -\bigl(\eta(x-i\beta)-\eta(x)\bigr)\bigl(\eta(x+i\beta)-\eta(x)\bigr),\\
  R_{-1}^{\text{dual}}\bigl(\eta(x)\bigr)&
  =\varepsilon\bigl(V_+(x)+V_-(x)\bigr)R_0^{\text{dual}}(\eta(x)).
  \label{dualclcon3}
\end{align}
By using the defining properties of the sinusoidal coordinate
\eqref{eta1+etam1}--\eqref{eta1*etam1}, we actually find that
$R_1^{\text{dual}}(z)$ is a degree 1 polynomial in $z$ and
$R_0^{\text{dual}}(z)$ is a quadratic polynomial:
\begin{align}
  R_1^{\text{dual}}(z)&=
  r_1^{(1)}z+r_{-1}^{(2)},
  \label{dualclR1}\\
  R_0^{\text{dual}}(z)&=
  r_1^{(1)}z^2+2r_{-1}^{(2)}z-\eta(-i\beta)\eta(i\beta).
  \label{dualclR0}
\end{align}
By using the explicit forms of $V_\pm$ \eqref{V+-=V+-t/etaeta}--\eqref{V+-t}
(with an arbitrary $L$) we obtain
\begin{equation}
  R_{-1}^{\text{dual}}(z)=\varepsilon\Bigl(
  v_{0,0}+\sum_{k=1}^L(v_{k,0}+v_{k-1,1})z^k\Bigr).
  \label{dualclR-1}
\end{equation}
Therefore all $R_i^{\text{dual}}(z)$ are polynomials and the dual
closure relation is demonstrated in a unified fashion for an arbitrary $L$.
Thus it does not characterise the exact nor the quasi-exact solvability.
For the exactly solvable $L=2$ case, by using \eqref{r10} and \eqref{rm11},
$R_{-1}^{\text{dual}}(z)$ can be written as
\begin{equation}
  R_{-1}^{\text{dual}}(z)=
  r_1^{(0)}z^2+r_{-1}^{(1)}z+\varepsilon v_{0,0}.
\end{equation}
For the real shifts case, in order to satisfy $D(0)=0$ \eqref{D(0)=0},
we have to take $v_{0,0}=-\eta(-1)v_{0,1}$ and this implies
$\varepsilon v_{0,0}=\eta(1)\eta(-1)B(0)$.
See (4.104)--(4.106) in \cite{os12}.

\subsection{Askey-Wilson algebra}
\label{AW3}

Here we will focus on the exactly solvable systems and will briefly
comment on the relationship between the closure plus the dual closure
relations and the so-called Askey-Wilson algebra
\cite{zhedanov,GLZ,vinzhed,terw}.
By simply expanding the double commutators in the closure
\eqref{closurerel} and the dual closure \eqref{dualclosurerel}
relations, we obtain two cubic relations generated by the two operators
$\mathcal{H}$ and $\eta$:
\begin{align}
  &\mathcal{H}^2\eta-(2+r_1^{(1)})\mathcal{H}\eta\mathcal{H}
  +\eta\mathcal{H}^2-r_1^{(0)}(\mathcal{H}\eta+\eta\mathcal{H})
  -r_0^{(0)}\eta
  =r_{-1}^{(2)}\mathcal{H}^2+r_{-1}^{(1)}\mathcal{H}+r_{-1}^{(0)},
  \label{expandclos}\\
  &\eta^2\mathcal{H}-(2+r_1^{(1)})\eta\mathcal{H}\eta+\mathcal{H}\eta^2
  -r_{-1}^{(2)}(\eta\mathcal{H}+\mathcal{H}\eta)
  +\eta(-i\beta)\eta(i\beta)\mathcal{H}
  =r_1^{(0)}\eta^2+r_{-1}^{(1)}\eta+\varepsilon v_{0,0}.
  \label{expanddualclos}
\end{align}
{}From its structure, the closure relation is at most linear in $\eta$
and at most quadratic in $\mathcal{H}$.
So the l.h.s. of \eqref{expandclos} has terms containing one factor
of $\eta$ and the r.h.s, none. It is simply $R_{-1}(\mathcal{H})$.
Likewise, the l.h.s. of \eqref{expanddualclos} has terms containing
one factor of $\mathcal{H}$ and the r.h.s, none. It is simply
$R^{\text{dual}}_{-1}(\mathcal{H})$. In \eqref{expanddualclos},
$\eta(-i\beta)\eta(i\beta)$ is just a real number, not an operator.

These have the same form as the so-called Askey-Wilson algebra, which has
many different expressions. 
The original one is due to Zhedanov \cite{zhedanov}.
Here we present a slightly more general version than the original one
and is due to \cite{GLZ, vinzhed}. It is generated by three elements
$K_1$, $K_2$, $K_3$:
\begin{align}
  [K_1,K_2]&=K_3,
  \label{glz1}\\
  [K_3,K_1]&=2\rho K_1K_2K_1+a_2(K_1K_2+K_2K_1)+a_1K_1^2
  +c_2K_2+dK_1+g_2,
   \label{glz2}\\
  [K_2,K_3]&=2\rho K_2K_1K_2+a_1(K_2K_1+K_1K_2)+a_2K_2^2
  +c_1K_1+dK_2+g_1.
   \label{glz3}
\end{align}
By expanding the commutators and eliminating $K_3$, they are reduced to
\begin{align}
  \!\!\!\!\!\!
  K_1^2K_2+2(\rho-1)K_1K_2K_1+K_2K_1^2+a_2(K_1K_2+K_2K_1)+c_2K_2
  &=-a_1K_1^2-dK_1-g_2,\\
  \!\!\!\!\!\!
  K_2^2K_1+2(\rho-1)K_2K_1K_2+K_1K_2^2+a_1(K_2K_1+K_1K_2)+c_1K_1
  &=-a_2K_2^2-dK_2-g_1.
\end{align}
Another version due to Terwilliger \cite{terw} is generated by two
independent elements $A$ and $A^{\times}$ and it has only expanded forms:
\begin{align}
  A^2A^{\times}-\beta_{\text{T}}AA^{\times}A+A^{\times}A^2
  -\gamma(AA^{\times}+A^{\times}A)-\rho_{\text{T}} A^{\times}
  &=\gamma^{\times}A^2+\omega A+\eta_{\text{T}},
  \label{AWrel1}\\
  A^{\times\,2}A-\beta_{\text{T}}A^{\times}AA^{\times}+AA^{\times\,2}
  -\gamma^{\times}(A^{\times}A+AA^{\times})-\rho_{\text{T}}^{\times}A
  &=\gamma A^{\times\,2}+\omega A^{\times}
  +\eta_{\text{T}}^{\times}.
  \label{AWrel2}
\end{align}
Here is the list of correspondence of the generators and coefficients:
\begin{equation}
  \begin{array}{ccccc}
  \text{ref.\,\cite{GLZ,vinzhed}}&\quad&\text{ref.\,\cite{terw}}
  &\quad&\text{this paper}\\
  K_1&&A&&\mathcal{H}\\
  K_2&&A^{\times}&&\eta\\
  2(1-\rho)&&\beta_{\text{T}}&&2+r_1^{(1)}\\
  -a_2&&\gamma&&r_1^{(0)}=\varepsilon(v_{2,0}+v_{1,1})\\
  -a_1&&\gamma^{\times}&&r_{-1}^{(2)}=\eta(-i\beta)+\eta(i\beta)\\
  -c_2&&\rho_{\text{T}}&&r_0^{(0)}=-\varepsilon^2v_{2,0}v_{1,1}\\
  -c_1&&\rho_{\text{T}}^{\times}&&-\eta(-i\beta)\eta(i\beta)\\
  -d&&\omega&&r_{-1}^{(1)}=\varepsilon(v_{1,0}+v_{0,1})\\
  -g_2&&\eta_{\text{T}}&&r_{-1}^{(0)}=-\varepsilon^2v_{2,0}v_{0,1}\\
  -g_1&&\eta_{\text{T}}^{\times}&&\varepsilon v_{0,0}.
  \end{array}
  \label{taiou}
\end{equation}
In \cite{GLZ} the Casimir operator $Q$  commuting with all the generators
of the algebra, $[K_1,Q]=[K_2,Q]=[K_3,Q]=0$ is given:
\begin{align}
  Q&=K_1K_2K_1K_2+K_2K_1K_2K_1
  -(1-\rho)(K_1K_2^2K_1+K_2K_1^2K_2)\n
  &\quad
  +(2-\rho)(a_1K_1K_2K_1+a_2K_2K_1K_2)
  +(1-\rho)(c_1K_1^2+c_2K_2^2)\\
  &\quad
  +(d-a_1a_2)(K_1K_2+K_2K_1)
  +\bigl((2-\rho)g_1-a_2c_1\bigr)K_1
  +\bigl((2-\rho)g_2-a_1c_2\bigr)K_2.
  \nonumber
\end{align}
With the above substitution \eqref{taiou}, $K_1\to\mathcal{H}$,
$K_2\to\eta$, etc, the Casimir operator turns out to be a constant
\cite{koornw}:
\begin{equation}
  Q=\varepsilon^2(v_{1,1}v_{0,0}-v_{1,0}v_{0,1}
  -r_{-1}^{(2)}v_{2,0}v_{0,1}).
\end{equation}
Although this fact might appear striking from the pure algebra point
of view \eqref{glz1}--\eqref{glz3}, it is rather trivial in quantum
mechanics.
In one-dimensional quantum mechanics, there is no dynamical operator
which commutes with the Hamiltonian.
Therefore, if $Q$ commutes with $\mathcal{H}$, it must be a constant.

Now here are some comments on the dissimilarity.
The first obvious difference is the structure. While the Askey-Wilson
algebra \eqref{glz1}--\eqref{glz3} or \eqref{AWrel1}--\eqref{AWrel2}
has no inherent structure, the closure relation \eqref{closurerel} has
the right structure to lead to the Heisenberg operator solution for
$\eta(x)$, whose positive and negative energy parts are the
annihilation-creation operators \cite{os7,os12,os13}.
It is the Hamiltonian and the annihilation-creation operators that
form the {\em dynamical symmetry algebra\/} of the system
\cite{os12,os13}, not the closure or dual-closure relations,
nor the Askey-Wilson algebra relations.
The $q$-oscillator algebra of \cite{os11} is the typical example
of the dynamical symmetry algebra thus obtained.

The next is the difference in character of the Askey-Wilson algebra
itself for the two cases; the pure imaginary shifts and the real
shifts cases.
The main scene of application of the Askey-Wilson algebra is the
theory of the orthogonal polynomials of a discrete variable.
The ($q$-)Racah polynomials are the typical example of this group
\cite{koeswart,os12}. In our language, it is the theory of the
eigenpolynomials of $\widetilde{\mathcal{H}}$ in discrete quantum
mechanics with real shifts. One outstanding feature of these polynomials
is the {\em duality\/} \cite{os12,terw}. For the eigenpolynomials
of $\widetilde{\mathcal{H}}$
\begin{equation}
  \widetilde{\mathcal{H}}P_n(\eta(x))=\mathcal{E}(n)P_n(\eta(x)),\quad
  n=0,1,\ldots,
\end{equation}
there exist the dual polynomials $Q_x(\mathcal{E}(n))$, satisfying
the relation
\begin{equation}
  P_n(\eta(x))=Q_x(\mathcal{E}(n)),\quad x=0,1,\ldots,\quad
  n=0,1,\ldots.
\end{equation}
This duality $x\leftrightarrow n$,
$\eta\sim\eta(x)\leftrightarrow \mathcal{E}(n)\sim\mathcal{H}$ is
reflected in the symmetry between the pair of operators (called the
Leonard pair \cite{leonard}) $K_1$ and $K_2$ or $A$ and $A^{\times}$
in the Askey-Wilson algebra.
The Askey-Wilson algebra or the closure and dual closure relations are
quite instrumental in clarifying various properties of the pair of
orthogonal polynomials of a discrete variable \cite{os12,terw}.

Now let us consider the discrete quantum mechanics with the pure
imaginary shifts.
In this case, the sinusoidal coordinate $\eta(x)$ takes the continuous
value (spectrum) for the continuous range of $x\in(x_1,x_2)$
\eqref{xconti}, which is markedly different from the spectrum of
$\mathcal{H}$ postulated to take the semi-infinite discrete values
\eqref{positivesemi}. The eigenpolynomials
$\widetilde{\mathcal{H}}P_n(\eta(x))=\mathcal{E}(n)P_n(\eta(x))$ depend
on the continuous parameter $x$ and they have no dual polynomials.
The Askey-Wilson and the Wilson polynomials are the typical examples
\cite{koeswart,os13}. As shown in previous work \cite{os7,os13} and in
\S\ref{sec:dualclosurerelation}, the essential information on exact
solvability is contained only in the closure relation \eqref{closurerel}.
There is no evidence that the dual closure relation plays a comparable
role to the closure relation. Therefore we may conclude that the
apparent symmetry between $\mathcal{H}$ and $\eta$, or $K_1$ and $K_2$
or $A$ and $A^{\times}$ in the Askey-Wilson algebra is quite misleading
for the pure imaginary shifts case.
In other words, a part of the Askey-Wilson algebra is irrelevant to
the orthogonal polynomials of a continuous variable.

\subsection{shape invariance}
\label{sec:shapeinvariance}

Let us briefly review the condition and the outcome of the shape
invariance \cite{genden} in our language.
In many cases the Hamiltonian contains some parameter(s),
$\bm{\lambda}=(\lambda_1,\lambda_2,\ldots)$.
Here we write parameter dependence explicitly, $\mathcal{H}(\bm{\lambda})$,
$\mathcal{A}(\bm{\lambda})$, $\mathcal{E}(n\,;\bm{\lambda})$,
$\phi_n(x\,;\bm{\lambda})$, etc, since it is the central issue.
The shape invariance condition is \cite{os4,os5,os12,os13}
\begin{equation}
  \mathcal{A}(\bm{\lambda})\mathcal{A}(\bm{\lambda})^{\dagger}
  =\kappa\mathcal{A}(\bm{\lambda'})^{\dagger}
  \mathcal{A}(\bm{\lambda'})
  +\mathcal{E}(1\,;\bm{\lambda}),
  \label{shapeinv}
\end{equation}
where $\kappa$ is a real positive parameter and $\bm{\lambda}'$ is
uniquely determined by $\bm{\lambda}$. Let us write the mapping as a function,
$\bm{\lambda}'=\text{si}(\bm{\lambda})$.
In concrete examples, if we take $\bm{\lambda}$ appropriately,
$\bm{\lambda}'$ has a simple additive form
$\bm{\lambda}'=\bm{\lambda}+\bm{\delta}$.
The energy spectrum and the excited state wavefunction are determined
by the data of the groundstate wavefunction $\phi_0(x\;\bm{\lambda})$
and the energy of the first excited state $\mathcal{E}(1\,;\bm{\lambda})$
as follows:
\begin{align}
  &\mathcal{E}(n\,;\bm{\lambda})=\sum_{s=0}^{n-1}
  \kappa^s\mathcal{E}(1\,;\bm{\lambda}^{[s]}),
  \label{shapeenery}\\
  &\phi_n(x\,;\bm{\lambda})\propto
  \mathcal{A}(\bm{\lambda}^{[0]})^{\dagger}
  \mathcal{A}(\bm{\lambda}^{[1]})^{\dagger}
  \mathcal{A}(\bm{\lambda}^{[2]})^{\dagger}
  \cdots
  \mathcal{A}(\bm{\lambda}^{[n-1]})^{\dagger}
  \phi_0(x\,;\bm{\lambda}^{[n]}).
  \label{phin=A..Aphi0}
\end{align}
Here $\bm{\lambda}^{[n]}$ is $\bm{\lambda}^{[0]}=\bm{\lambda}$,
$\bm{\lambda}^{[n]}=\text{si}(\bm{\lambda}^{[n-1]})$ ($n=1,2,\ldots$).

\subsubsection{pure imaginary shifts case}
\label{sec:shapeinv_cont}

Here is a unified proof of the shape invariance for the discrete
quantum mechanics with pure imaginary shifts.
The shape invariance condition \eqref{shapeinv} is decomposed to
the following set of two equations:
\begin{align}
  &V(x-i\tfrac{\gamma}{2}\,;\bm{\lambda})
  V^*(x-i\tfrac{\gamma}{2}\,;\bm{\lambda})
  =\kappa^2\,V(x\,;\bm{\lambda}')V^*(x-i\gamma\,;\bm{\lambda}'),
  \label{contshape1}\\
  &V(x+i\tfrac{\gamma}{2}\,;\bm{\lambda})
  +V^*(x-i\tfrac{\gamma}{2}\,;\bm{\lambda})
  =\kappa\bigl(V(x\,;\bm{\lambda}')+V^*(x\,;\bm{\lambda}'))
  -\mathcal{E}(1\,;\bm{\lambda}).
  \label{contshape2}
\end{align}

We assume that $\eta(x)$ satisfies the relation
\begin{equation}
  \eta(x)=[\tfrac12]\bigl(
  \eta(x-i\tfrac{\gamma}{2})+\eta(x+i\tfrac{\gamma}{2})
  -\eta(-i\tfrac{\gamma}{2})-\eta(i\tfrac{\gamma}{2})\bigr).
  \label{etahalf}
\end{equation}
Moreover we assume that $\eta(x\,;\bm{\lambda}')=\eta(x\,;\bm{\lambda})$,
for example it is satisfied if $\eta(x)$ is $\bm{\lambda}$-independent.
Both are easily verified in each of the explicit examples listed in
the Appendix \ref{app:sinusoidal}, \eqref{etaform1}--\eqref{etaform8}.
When the forms of the potential functions \eqref{V+-=V+-t/etaeta} and
\eqref{V+-t} (with $L=2$) are substituted, the shape invariance
conditions \eqref{contshape1}--\eqref{contshape2} are satisfied.
If we take $\{v_{k,0}\ (k=0,1,2),v_{k,1}\ (k=0,1)\}$ as $\bm{\lambda}$,
then $\bm{\lambda'}$ and $\mathcal{E}(1\,;\bm{\lambda})$ are
\begin{align}
  \kappa v'_{2,0}&=-v_{1,1},
  \label{v20formula}\\
  \kappa v'_{1,1}&=v_{2,0}+[2]v_{1,1},
  \label{v11formula}\\
  \kappa v'_{1,0}&=[\tfrac12](v_{1,0}-v_{0,1})
  +r_{-1}^{(2)}\Bigl(\frac{[\frac14]^2}{[\frac12]}\,v_{2,0}
  +v_{1,1}\Bigr),
  \label{v10formula}\\
  \kappa v'_{0,1}&=[\tfrac12]v_{1,0}+[\tfrac32]v_{0,1}
  +r_{-1}^{(2)}\Bigl(\frac{[\frac14]^2}{[\frac12]}\,v_{2,0}
  +\frac{[\frac14][\frac34]}{[\frac12]^2}\,v_{1,1}\Bigr),\\
  \kappa v'_{0,0}&=v_{0,0}
  +r_{-1}^{(2)}\Bigl(\frac{[\frac14][\frac34]}{[\frac12]}\,v_{0,1}
  -\frac{[\frac14]^2}{[\frac12]}\,v_{1,0}\Bigr)
  +[\tfrac12]^2\Bigl(\frac{[\frac14]^4}{[\frac12]^4}\,r_{-1}^{(2)\,2}
  -\eta(-i\gamma)\eta(i\gamma)\Bigr)v_{2,0}\n
  & -[\tfrac12]\Bigl(\frac{[\frac14]^3[\frac34]}{[\frac12]^3}\,r_{-1}^{(2)\,2}
  +[\tfrac32]\eta(-i\gamma)\eta(i\gamma)\Bigr)v_{1,1},
  \label{v00formula}\\
  \mathcal{E}(1\,;\bm{\lambda})&=v_{1,1}.
  \label{e1formula}
\end{align}
Note that the above formula $\mathcal{E}(1\,;\bm{\lambda})$ is
consistent with the general formula \eqref{energyeigen}. 
It is elementary to verify that the quadratic recursion formula
generated by \eqref{v20formula} and \eqref{v11formula} coupled with
the shape invariance energy formulas \eqref{shapeenery} and
\eqref{e1formula} reproduces the energy eigenvalue formula
\eqref{energyeigen}.
However, the other formulas for the parameter shifts
\eqref{v10formula}--\eqref{v00formula} seem too complicated to be practical.
As shown in \cite{os4,os5,os7,os13}, the parameter shifts are much
simpler for the known examples.
 
\paragraph{remark}

In ordinary quantum mechanics there is a method for constructing
a family of isospectral Hamiltonians, known as Crum's theorem \cite{crum}.
Recently we have obtained its discrete quantum mechanics version,
see \cite{os15}.
If $\phi_1(x)$ take a form
$\phi_1(x)=\phi_0(x)(\text{const}+\text{const}\cdot\eta(x))$,
which occurs indeed in the setting of this paper,
the potential function of the first associated Hamiltonian is given by
\begin{equation}
  V^{[1]}(x+i\tfrac{\gamma}{2})
  =V(x)\,\frac{\eta(x-i\gamma)-\eta(x)}{\eta(x)-\eta(x+i\gamma)}\,.
\end{equation}
Therefore, if shape invariance holds, $V(x)$ satisfies
\begin{equation}
  V(x+i\tfrac{\gamma}{2}\,;\bm{\lambda}')
  =\kappa^{-1}\,V(x\,;\bm{\lambda})\,
  \frac{\eta(x-i\gamma\,;\bm{\lambda})-\eta(x\,;\bm{\lambda})}
  {\eta(x\,;\bm{\lambda})-\eta(x+i\gamma\,;\bm{\lambda})},
\end{equation}
in which the sinusoidal coordinate may depend on $\bm{\lambda}$.

\subsubsection{real shifts case}
\label{sec:shapeinv_dics}

The shape invariance \eqref{shapeinv} is equivalent to the following
set of two equations:
\begin{align}
  &B(x+1\,;\bm{\lambda})D(x+1\,;\bm{\lambda})
  =\kappa^2\,B(x\,;\bm{\lambda}')D(x+1\,;\bm{\lambda}'),\\
  &B(x\,;\bm{\lambda})+D(x+1\,;\bm{\lambda})
  =\kappa\bigl(B(x\,;\bm{\lambda}')+D(x\,;\bm{\lambda}'))
  +\mathcal{E}(1\,;\bm{\lambda}).
\end{align}
For the classified five types of $\eta(x)$,
$\text{(\romannumeral1)}'$--$\text{(\romannumeral5)}'$
in \eqref{etaform1'}--\eqref{etaform5'}, the shape invariance holds.
The boundary condition $D(0)=0$ \eqref{D(0)=0} forces to
choose $v_{0,0}$ as $v_{0,0}=-v_{0,1}\eta(-1)$. Thus we take the
parameters $\{v_{k,0}\ (k=1,2),v_{k,1}\ (k=0,1)\}$
(and $d$ for $\text{(\romannumeral2)}'$ and $\text{(\romannumeral5)}'$)
as $\bm{\lambda}$, then $\bm{\lambda'}$ and
$\mathcal{E}(1\,;\bm{\lambda})$ are
\begin{align}
  \kappa v'_{2,0}&=-v_{1,1},
  \label{v20formuladisc}\\
  \kappa v'_{1,1}&=v_{2,0}+[2]v_{1,1},
  \label{v11formuladisc}\\
  \kappa v'_{1,0}&=\mu[\tfrac12](v_{1,0}-v_{0,1})
  +\mu[\tfrac12]\eta(1)v_{2,0}+\nu r_{-1}^{[2]}v_{1,1},\\
  \kappa v'_{0,1}&=\mu\bigl([\tfrac12]v_{1,0}+[\tfrac32]v_{0,1}\bigr)
  +\mu[\tfrac12]\eta(1)v_{2,0}
  +\Bigl(\nu r_{-1}^{[2]}+\mu[\tfrac12]\bigl(\eta(1)-\eta(-1)\bigr)\Bigr)
  v_{1,1},\\
  \mathcal{E}(1\,;\bm{\lambda})&=-v_{1,1},
  \label{e1formuladisc}\\
  d'&=
  \begin{cases}
   d+1&\text{for $\text{(\romannumeral2)}'$}\\
   d q&\text{for $\text{(\romannumeral5)}'$},
  \end{cases}
\end{align}
in which $\mu$ and $\nu$ are constants
\begin{equation}
  \mu=
  \begin{cases}
   1&\text{for $\text{(\romannumeral1)}'$--$\text{(\romannumeral2)}'$}\\
   q^{-\frac12}&\text{for $\text{(\romannumeral3)}'$}\\
   q^{\frac12}&\text{for $\text{(\romannumeral4)}'$--
   $\text{(\romannumeral5)}'$},
  \end{cases}
  \qquad
  \nu=
  \begin{cases}
   1&\text{for $\text{(\romannumeral1)}'$--$\text{(\romannumeral4)}'$}\\
   {\displaystyle\frac{1+dq}{1+d}}&\text{for $\text{(\romannumeral5)}'$}.
  \end{cases}
\end{equation}
The quadratic recursion formula \eqref{v20formuladisc} and
\eqref{v11formuladisc} are exactly the same as those of the pure
imaginary shifts case \eqref{v20formula} and \eqref{v11formula}.
Therefore the shape invariance energy formulas \eqref{shapeenery} and
\eqref{e1formuladisc} produce the same energy spectra \eqref{energyeigen}.
For the finite dimensional case, the natural number $N$ satisfying
$B(N\,;\bm{\lambda})=0$ \eqref{D(0)=0} is also counted as a varying
parameter. Then the shape invariance including the conditions
\begin{equation}
  N'=N-1,\quad B(N'\,;\bm{\lambda}')=0
\end{equation}
is satisfied.
As shown in \cite{os12}, the parameter shifts are much
simpler for the known examples.
 
\section{Quasi-Exactly Solvable $\widetilde{\mathcal{H}}'$}
\label{sec:QES}
\setcounter{equation}{0}

Quasi-exact solvability (QES) means that only a finite part of the
spectrum and the corresponding eigenfunctions can be obtained exactly
\cite{Ush}.
Usually such a theory contains a finite dimensional vector space
\cite{turb} consisting of polynomials of a certain degree which forms
an invariant subspace of the `Hamiltonian' $\widetilde{\mathcal{H}}$,
or more precisely its modification $\widetilde{\mathcal{H}}'$.
There are many ways to accomplish QES. The method of this paper can be
considered as a simple generalisation of the one in \cite{st1}.
That is, to add non-solvable higher order term(s) together with
compensation term(s) to an exactly solvable theory.
As is clear from the construction, the sinusoidal coordinate plays
an essential role.

For a given positive integer $M$, let us try to find a QES `Hamiltonian'
$\widetilde{\mathcal{H}}$, or more precisely its modification
$\widetilde{\mathcal{H}}'$, having an invariant polynomial space
$\mathcal{V}_M$:
\begin{equation}
  \widetilde{\mathcal{H}}'\mathcal{V}_M\subseteq\mathcal{V}_M.
\end{equation}
For $L\geq 3$, \eqref{Htetan} is
\begin{equation}
  \widetilde{\mathcal{H}}\eta(x)^n
  =\sum_{m=0}^{L-3}\eta(x)^{n+L-2-m}\sum_{j=0}^me_{m,j,n}
  +\bigl(\text{a polynomial of degree $n$ in $\eta(x)$}\bigr).
\end{equation}
So let us define $\widetilde{\mathcal{H}}'$ by adding compensation
terms to $\widetilde{\mathcal{H}}$ as
\begin{equation}
  \widetilde{\mathcal{H}}'\eqdef
  \widetilde{\mathcal{H}}-\sum_{m=0}^{L-3}e_m(M)\eta(x)^{L-2-m},\quad
  e_m(M)\eqdef\sum_{j=0}^me_{m,j,M}.
\end{equation}
Then we have $\widetilde{\mathcal{H}}'\eta(x)^M\in\mathcal{V}_M$.
For $1\leq m'\leq L-3$, we have
\begin{align}
  \widetilde{\mathcal{H}}'\eta(x)^{M-m'}
  &=\sum_{m=0}^{L-m'-3}\eta(x)^{M+L-m'-2-m}\Bigl(
  \sum_{j=0}^me_{m,j,M-m'}-e_m(M)\Bigr)\n
  &\quad
  +\bigl(\text{a polynomial of degree $M$ in $\eta(x)$}\bigr).
\end{align}
If we could choose $v_{k,l}$ to satisfy all these conditions
\begin{equation}
  \sum_{j=0}^me_{m,j,M-m'}-e_m(M)=0\quad
  (1\leq m'\leq L-3,\ 0\leq m\leq L-m'-3),
  \label{qescondL}
\end{equation}
then we would obtain
$\widetilde{\mathcal{H}}'\mathcal{V}_M\subseteq\mathcal{V}_M$.

\subsection{QES with $L=3$ }
\label{sec:L=3}

For the $L=3$ case, $\widetilde{\mathcal{H}}'$ is defined by adding
one compensation term of degree one
\begin{equation}
  \widetilde{\mathcal{H}}'\eqdef\widetilde{\mathcal{H}}-e_0(M)\eta(x),\quad
  e_0(M)
  \eqdef e_{0,0,M},
  \label{L3ham}
\end{equation}
and we have achieved the quasi-exact solvability
$\widetilde{\mathcal{H}}'\mathcal{V}_M\subseteq\mathcal{V}_M$.
The number of exactly determined eigenstates is
$M+1=\text{dim}(\mathcal{V}_M)$.
In this case there is no extra conditions for $v_{k,l}$.
The explicit form of $e_0(M)$ is
\begin{equation}
  e_0(M)=\varepsilon\frac{[\frac{M}{2}]}{[\frac12]}
  \bigl([\tfrac{M-1}{2}]v_{3,0}+[\tfrac{M+1}{2}]v_{2,1}\bigr).
  \label{e0M}
\end{equation}
This QES theory has two more parameters $v_{3,0}$ and $v_{2,1}$
on top of those in the original exactly-solvable theory ($L=2$).
Most known examples of QES belong to this category but those in
ordinary quantum mechanics have only one extra parameter.

\subsection{QES with $L=4$}
\label{sec:L=4}

This type of QES theory is new.
For $L=4$ case, $\widetilde{\mathcal{H}}'$ is defined by adding a
linear and a quadratic in $\eta(x)$ compensation terms to the
Hamiltonian $\widetilde{\mathcal{H}}$:
\begin{equation}
  \widetilde{\mathcal{H}}'\eqdef
  \widetilde{\mathcal{H}}-e_0(M)\eta(x)^2-e_1(M)\eta(x),\quad
  e_0(M)\eqdef e_{0,0,M},\quad
  e_1(M)\eqdef e_{1,0,M}+e_{1,1,M},
  \label{L4ham}
\end{equation}
and $\widetilde{\mathcal{H}}'\eta(x)^M\in\mathcal{V}_M$.
By using \eqref{[a][a+c]..} we have
\begin{align}
  &\widetilde{\mathcal{H}}'\eta(x)^{M-1}
  =\eta(x)^{M+1}\bigl(e_{0,0,M-1}-e_0(M)\bigr)
  +\bigl(\text{a polynomial of degree $M$ in $\eta(x)$}\bigr)\n
  &\qquad\ \ 
  =-\varepsilon\eta(x)^{M+1}\bigl([M-1]v_{4,0}+[M]v_{3,1}\bigr)
  +\bigl(\text{a polynomial of degree $M$ in $\eta(x)$}\bigr).
  \label{L4HetaM-1}
\end{align}
In order to eliminate the $\eta(x)^{M+1}$ term, we choose $v_{3,1}$ as
\begin{equation}
  v_{3,1}=-\frac{[M-1]}{[M]}\,v_{4,0}.
  \label{v31}
\end{equation}
We have achieved the quasi-exact solvability
$\widetilde{\mathcal{H}}'V_M\subseteq V_M$.
The explicit forms of $e_0(M)$ and $e_1(M)$ are
\begin{align}
  e_0(M)&=-\varepsilon
  \frac{[4][\frac{M}{2}][\frac{M-1}{2}]}{[\frac12][M+3]}\,v_{4,0},
  \label{L4e0M}\\
  e_1(M)&=\varepsilon\frac{[\frac{M}{2}]}{[\frac12]}
  \bigl([\tfrac{M-1}{2}]v_{3,0}+[\tfrac{M+1}{2}]v_{2,1}\bigr)\n
  &\quad -\varepsilon r_{-1}^{(2)}v_{4,0}\times
  \begin{cases}
    \frac{M(M-1)(M^2+5M+8)}{M+3}
    &\text{for }r_1^{(1)}=0,\\
    \frac{2[\frac{M}{2}][\frac{M-1}{2}]}{r_1^{(1)}[\frac12][M+3]}
    \Bigl([4]-2[3]+2[\frac12]\frac{[2M+5]}{[\frac{2M+5}{2}]}\Bigr)
    &\text{for }r_1^{(1)}\neq 0.
  \end{cases}
  \label{L4e1M}
\end{align}
The theory has three more free parameters on top of those of the
original exactly solvable theory ($L=2$).

\subsection{non-QES for $L\geq 5$}
\label{sec:L>=5}

The higher $L$ becomes, the number of conditions to be satisfied
\eqref{qescondL} increases more rapidly than the number of additional
parameters.
We will show that $L\geq 5$ case cannot be made QES.
The condition \eqref{qescondL} with $m=0$ gives
$e_{0,0,M}=e_{0,0,M-m'}$ ($1\leq m'\leq L-3$),
and by using \eqref{e_{m,0,n}} and \eqref{[a][a+c]..} we obtain
\begin{equation}
  [M-\tfrac{m'+1}{2}]v_{L,0}+[M-\tfrac{m'-1}{2}]v_{L-1,1}=0\quad
  (1\leq m'\leq L-3).
\end{equation}
For $L\geq 5$ case, these equations do not have non-trivial solutions.
For $m'=1,2$ we obtain
\begin{equation}
  \begin{pmatrix}
    [M-1]&[M]\\
    [M-\frac32]&[M-\frac12]
  \end{pmatrix}
  \begin{pmatrix}v_{L,0}\\v_{L-1,1}\end{pmatrix}
  =\begin{pmatrix}0\\0\end{pmatrix}.
\end{equation}
The determinant of this matrix is $[\frac12]$ which does not vanish.
Thus we obtain $v_{L,0}=v_{L-1,1}=0$.
Namely there is no $v_{k,l}$ ($k+l=L$) term.
Therefore $L\geq 5$ case cannot be made QES.

\section{(Quasi-)Exactly Solvable Hamiltonian}
\label{sec:QESH}
\setcounter{equation}{0}

If there exists a groundstate wavefunction $\phi_0(x)$ which
satisfies \eqref{Aphi0=0} (and $|\!|\phi_0|\!|<\infty$, the hermiticity
of $\mathcal{H}$), we can return to the Hamiltonian $\mathcal{H}$ from
the `Hamiltonian' $\widetilde{\mathcal{H}}$ by the inverse similarity
transformation \eqref{Htdef}.
In the same way the QES Hamiltonian $\mathcal{H}'$ is obtained from
$\widetilde{\mathcal{H}}'$ by the inverse similarity transformation
in terms of the {\em pseudo-groundstate\/} wavefunction $\phi_0(x)$
satisfying $\mathcal{A}\phi_0=0$ \eqref{Aphi0=0},
\begin{equation}
  \mathcal{H}'\eqdef
  \phi_0(x)\circ\widetilde{\mathcal{H}}'\circ\phi_0(x)^{-1}.
  \label{invsimtr}
\end{equation}
It should be noted that $\phi_0(x)$ is neither the groundstate nor
an eigenstate of the total Hamiltonian $\widetilde{\mathcal{H}}'$.
Thus it is called the pseudo-groundstate wavefunction.
For the $L=3,4$ cases we have
\begin{align}
  L=3\ :&\quad
  \mathcal{H}'\eqdef\mathcal{H}-e_0(M)\eta(x),\\
  L=4\ :&\quad
  \mathcal{H}'\eqdef\mathcal{H}-e_0(M)\eta(x)^2-e_1(M)\eta(x).
\end{align}
Let us note that the Hamiltonian $ \mathcal{H}'$ does not factorise and
the semi positive-definite spectrum is lost due to the compensation terms.
For the pure imaginary shifts case, the existence of (pseudo-)groundstate
wavefunction $\phi_0(x)$ strongly depends on the concrete form of
$V(x)$ and its parameter range. There is no general formula to write
down $\phi_0(x)$ in terms of $V(x)$.
On the other hand, for the real shifts case, the (pseudo-)groundstate
wavefunction $\phi_0(x)$ is uniquely given by \cite{os12}
\begin{equation}
  \phi_0(x)=\sqrt{\prod_{y=0}^{x-1}\frac{B(y)}{D(y+1)}}.
  \label{dicsrete_phi0}
\end{equation}
The positivity of $B(x)$ and $D(x)$ \eqref{D(0)=0} restricts their
parameter range. For the infinite case $x\in[0,\infty)$,
the square-summability $|\!|\phi_0|\!|<\infty$ restricts the asymptotic
forms of $B(x)$ and $D(x)$.

\section{Summary and the Recipe}
\label{sec:summary}
\setcounter{equation}{0}

Based on the sinusoidal coordinate $\eta(x)$ we have systematically
explored a unified theory of one-dimensional exactly and quasi-exactly
solvable `discrete' quantum mechanical models.
The Hamiltonians of discrete quantum mechanics have shift operators
as exponentiated forms of the momentum operator $p=-i\partial_x$,
$e^{\pm \beta p}=e^{\mp i\beta\partial_x}$.
This method applies to both the pure imaginary shifts
($\beta=\gamma\in\mathbb{R}_{\neq0}$) and the real shifts cases
($\beta=i=\sqrt{-1}$), which have a continuous and a discrete dynamical
variable $x$, respectively.
The main input is the special form of the potential functions $V_{\pm}(x)$
\eqref{V+-=V+-t/etaeta} and \eqref{V+-t}, with which the `Hamiltonian'
$\widetilde{\mathcal{H}}$ \eqref{Ht} maps a polynomial in $\eta(x)$ into
another.
We obtain exactly solvable models (degree $n\to n$) and quasi-exactly
solvable models (degree $n\to n+1,n+2$) by adding compensation terms,
which are linear and quadratic in $\eta(x)$, respectively.
The QES Hamiltonians based on the mapping (degree $n\to n+2$) are new.

The corresponding result in
the ordinary QM can be found in the Appendix of \cite{os7}.
This early work, however, does not cover the quasi-exactly solvable cases.
In this connection, see the recent developments \cite{ho}.

The present paper is for the presentation of the basic formalism.
Application and concrete examples will be explored in a subsequent
publication \cite{os16}. The explicit forms of various sinusoidal
coordinates are listed in Appendix \ref{app:sinusoidal}.

The forms of the (pseudo-)groundstate wavefunctions $\phi_0(x)$ for
the pure imaginary shifts (the continuous variable) case depend on the
choice of the sinusoidal coordinates.
They are `gamma functions' having various shift properties;
the (Euler) gamma function for (\romannumeral1)--(\romannumeral2)
\eqref{etaform1}--\eqref{etaform2},
the $q$-gamma function (or the $q$-Pochhammer symbol) for
(\romannumeral3)--(\romannumeral4) \eqref{etaform3}--\eqref{etaform4},
the double gamma function (or the quantum dilogarithm function) for
(\romannumeral5)--(\romannumeral8) \eqref{etaform5}--\eqref{etaform8}.
In a subsequent publication we will present explicit examples of
new Hamiltonians based on \eqref{etaform5}--\eqref{etaform8}.
Their eigenfunctions contain orthogonal polynomials and the double
gamma functions as the orthogonality measure functions.
Eigenpolynomials $P_n(\eta(x))$ for exactly solvable QM belong to
the Askey-scheme of hypergeometric orthogonal polynomials.
Various examples of exactly solvable QM 
were investigated for (\romannumeral1)--(\romannumeral4)
\eqref{etaform1}--\eqref{etaform4} \cite{os13} and
$\text{(\romannumeral1)}'$--$\text{(\romannumeral5)}'$
\eqref{etaform1'}--\eqref{etaform5'} \cite{os12},
and quasi-exactly solvable QM 
were partially examined in \cite{os10,newqes}. 

The simple recipe to construct an exactly or quasi-exactly solvable
Hamiltonian is as follows:
\begin{enumerate}
\item[(1)] Choose the sinusoidal coordinate among
\eqref{etaform1}--\eqref{etaform8} if the variable is continuous,
among \eqref{etaform1'}--\eqref{etaform5'} for the discrete variable.
\item[(2)] Choose $L=2$ for exact solvability and $L=3,4$ for
quasi-exact solvability and write down the `Hamiltonian'
$\widetilde{\mathcal{H}}$ in the polynomial space
\eqref{V+-=V+-t/etaeta} and \eqref{V+-t} with the free parameters
$v_{k,l}$, $k+l\le L$, $l=0,1$.
For the quasi-exactly solvable case, add the proper compensation terms
\eqref{L3ham}--\eqref{e0M} for the $L=3$ case and
\eqref{L4ham}--\eqref{L4e1M} for the $L=4$ case.
\item[(3)] Determine the (pseudo-)groundstate $\phi_0$ as a zero mode
of $\mathcal{A}$, \eqref{Aphi0=0}, which can be found among the
various gamma functions listed as above or \eqref{dicsrete_phi0}
for the discrete variable case.
\item[(4)] Restrict the parameter ranges so that the square-integrability
of $\phi_0$ and the hermiticity is satisfied for the continuous
variable case. For the discrete variable case the positivity $B(x)>0$
and $D(x)>0$ and the boundary condition(s) $D(0)=0$, ($B(N)=0$)
\eqref{D(0)=0} are the conditions to restrict the parameters.
For the infinite dimensional case the square summability
$\sum_{x=0}^\infty\phi_0^2(x)<\infty$ must be satisfied, too.
\item[(5)] Apply the inverse similarity transformation \eqref{invsimtr}
in terms of $\phi_0$ on the `Hamiltonian' in the polynomial space
to get the Hamiltonian $\mathcal{H}$ or $\mathcal{H}'$.
\end{enumerate}

\section*{Acknowledgements}
We thank Paul Terwilliger for fruitful discussion.
This work is supported in part by Grants-in-Aid for Scientific Research
from the Ministry of Education, Culture, Sports, Science and Technology,
No.18340061 and No.19540179.

\appendix
\section{List of Sinusoidal Coordinates}
\label{app:sinusoidal}
\setcounter{equation}{0}
\renewcommand{\theequation}{A.\arabic{equation}}

Here we list the explicit forms of the sinusoidal coordinates,
based on which various concrete examples of exactly and quasi-exactly
solvable theories are constructed.

These are eight sinusoidal coordinates for the pure imaginary shifts
case (continuous $x$):
\begin{alignat}{4}
  \text{(\romannumeral1)}:&\quad&\eta(x)&=x,
  &\quad -\infty<\,&x<\infty,&\quad&\gamma=1,
  \label{etaform1}\\
  \text{(\romannumeral2)}:&\quad&\eta(x)&=x^2,
  &\quad 0<\,&x<\infty,&\quad&\gamma=1,
  \label{etaform2}\\
  \text{(\romannumeral3)}:&\quad&\eta(x)&=1-\cos x,
  &\quad 0<\,&x<\pi,&\quad&\gamma\in\mathbb{R}_{\neq 0},
  \label{etaform3}\\
  \text{(\romannumeral4)}:&\quad&\eta(x)&=\sin x,
  &\quad -\tfrac{\pi}{2}<\,&x<\tfrac{\pi}{2},
  &\quad&\gamma\in\mathbb{R}_{\neq 0},
  \label{etaform4}\\
  \text{(\romannumeral5)}:&\quad&\eta(x)&=1-e^{-x},
  &\quad -\infty<\,&x<\infty,&\quad&\gamma\in\mathbb{R}_{\neq 0},
  \label{etaform5}\\
  \text{(\romannumeral6)}:&\quad&\eta(x)&=e^x-1,
  &\quad -\infty<\,&x<\infty,&\quad&\gamma\in\mathbb{R}_{\neq 0},
  \label{etaform6}\\
  \text{(\romannumeral7)}:&\quad&\eta(x)&=\cosh x-1,
  &\quad 0<\,&x<\infty,&\quad&\gamma\in\mathbb{R}_{\neq 0},
  \label{etaform7}\\
  \text{(\romannumeral8)}:&\quad&\eta(x)&=\sinh x,
  &\quad -\infty<\,&x<\infty, &\quad&\gamma\in\mathbb{R}_{\neq 0},
  \label{etaform8}
\end{alignat}
and five sinusoidal coordinates for the real shifts case (integer $x$):
\begin{alignat}{2}
  \text{(\romannumeral1)}':&\quad&\eta(x)&=x,
  \label{etaform1'}\\
  \text{(\romannumeral2)}':&\quad&\eta(x)&=\epsilon'x(x+d),
  \qquad \qquad \quad \
  \epsilon'=\Bigl\{\begin{array}{ll}
  1&\text{for }\ d>-1,\\[4pt]
  -1&\text{for }\ d<-N,
  \label{etaform2'}
  \end{array}\\
  \text{(\romannumeral3)}':&\quad&\eta(x)&=1-q^x,
  \label{etaform3'}\\
  \text{(\romannumeral4)}':&\quad&\eta(x)&=q^{-x}-1,
  \label{etaform4'}\\
  \text{(\romannumeral5)}':&\quad&\eta(x)&=\epsilon'(q^{-x}-1)(1-dq^x),\quad
  \epsilon'=\Bigl\{\begin{array}{ll}
  1&\text{for }\ d<q^{-1},\\[4pt]
  -1&\text{for }\ d>q^{-N},
  \end{array}
  \label{etaform5'}
\end{alignat}
where $0<q<1$.
As shown in detail in \S4C of \cite{os12}, the above five sinusoidal
coordinates for the real shifts \eqref{etaform1'}--\eqref{etaform5'}
exhaust all the solutions of \eqref{eta1+etam1}--\eqref{eta1*etam1}
up to a multiplicative factor. On the other hand, those for the pure
imaginary shifts (\romannumeral1)--(\romannumeral8)
\eqref{etaform1}--\eqref{etaform8} are merely typical examples
satisfying all the postulates for the sinusoidal coordinate
\eqref{eta1+etam1}--\eqref{eta1*etam1} and the extra one used for the
shape invariance \eqref{etahalf}.
It is easy to see that $\eta(x)=x+\sinh(2\pi x)$, $-\infty<x<\infty$,
$\gamma=1$ is a good sinusoidal coordinate for the imaginary shifts
but it fails to fulfill the extra condition \eqref{etahalf}.

\section{Hermiticity of the Hamiltonian}
\label{app:hermiticity}
\setcounter{equation}{0}
\renewcommand{\theequation}{B.\arabic{equation}}

In this Appendix we recapitulate the proof of the hermiticity of the
Hamiltonian $\mathcal{H}$ \eqref{H} for the pure imaginary shifts
(continuous variable) case \cite{os10,os13}.
For the real shifts (discrete variable) case the Hamiltonian
$\mathcal{H}$ \eqref{H} is a hermitian matrix (real symmetric matrix)
and there is no problem for the hermiticity.
Thus we consider only the continuous variables case, in which
the wavefunctions and the potential functions are analytic function
of $x$ as explained in \S\ref{discQM}.
The $*$-operation on analytic functions is also defined there.

By using the formula $(AB)^{\dagger}=B^{\dagger}A^{\dagger}$,
we obtain $\mathcal{H}^{\dagger}=\mathcal{H}$ but this is formal
hermiticity.
In order to demonstrate the true hermiticity of $\mathcal{H}$,
we have to show $(g,\mathcal{H}f)=(\mathcal{H}g,f)$ with respect to
the inner product \eqref{inn_pro}.
Since the eigenfunctions considered in this paper have the form
$\phi_0(x)P_n\bigl(\eta(x)\bigr)$, it is sufficient to check for
$f(x)=\phi_0(x)P\bigl(\eta(x)\bigr)$ and
$g(x)=\phi_0(x)Q\bigl(\eta(x)\bigr)$, where $P(\eta)$ and $Q(\eta)$ are
polynomials in $\eta$.
Since $\mathcal{H}$ is real $\mathcal{H}^*=\mathcal{H}$, namely
$\mathcal{H}$ maps a `real' function to a `real' function ($f^*(x)=f(x)
\Rightarrow(\mathcal{H}f)^*=(\mathcal{H}f)$), we can take
$\phi_0(x)$, $P\bigl(\eta(x)\bigr)$ and $Q\bigl(\eta(x)\bigr)$ to be
`real' functions of $x$ and we do so in the following.

The (pseudo-)groundstate wavefunction $\phi_0(x)$ is determined as
a zero mode of $\mathcal{A}$, $\mathcal{A}\phi_0=0$  \eqref{Aphi0=0}.
The equation reads
\begin{equation}
  \sqrt{V^*(x-i\tfrac{\gamma}{2})}\,\phi_0(x-i\tfrac{\gamma}{2})
  =\sqrt{V(x+i\tfrac{\gamma}{2})}\,\phi_0(x+i\tfrac{\gamma}{2}).
  \label{explicit_Aphi0=0}
\end{equation}
Let us define $T_+=\sqrt{V(x)}\,e^{\gamma p}\sqrt{V^*(x)}$ and
$T_-=\sqrt{V^*(x)}\,e^{-\gamma p}\sqrt{V(x)}$.
Then the Hamiltonian \eqref{H} is $\mathcal{H}=T_++T_--V(x)-V^*(x)$.
For the QES case, the compensation terms are added.
It is obvious that the function part $-V(x)-V^*(x)$ (plus possible
compensation terms) is hermitian by itself.
Let us define two analytic functions $F(x)$ and $G(x)$ as follows:
\begin{align}
  \!
  g^*(x)T_+f(x)&=\phi^*_0(x)Q^*\bigl(\eta^*(x)\bigr)
  \sqrt{V(x)}\sqrt{V^*(x-i\gamma)}\,
  \phi_0(x-i\gamma)P\bigl(\eta(x-i\gamma)\bigr)
  \eqdef F(x),\\
  \!
  g^*(x)T_-f(x)&=\phi^*_0(x)Q^*\bigl(\eta^*(x)\bigr)
  \sqrt{V^*(x)}\sqrt{V(x+i\gamma)}\,
  \phi_0(x+i\gamma)P\bigl(\eta(x+i\gamma)\bigr)
  \eqdef G(x).
\end{align}
Then we have
\begin{align}
  \!
  (T_+g)^*(x)f(x)&=\phi^*_0(x+i\gamma)Q^*\bigl(\eta^*(x+i\gamma)\bigr)
  \sqrt{V^*(x)}\sqrt{V(x+i\gamma)}\,\phi_0(x)P\bigl(\eta(x)\bigr)
  =F(x+i\gamma),
  \label{(T+g)^*f)}\\
  \!
  (T_-g)^*(x)f(x)&=\phi^*_0(x-i\gamma)Q^*\bigl(\eta^*(x-i\gamma)\bigr)
  \sqrt{V(x)}\sqrt{V^*(x-i\gamma)}\,\phi_0(x)P\bigl(\eta(x)\bigr)
  =G(x-i\gamma).
  \label{(T-g)^*f)}
\end{align}
By using \eqref{explicit_Aphi0=0} and the `reality' of $\phi_0(x)$,
$\eta(x)$, $P(\eta)$, $Q(\eta)$, we obtain
\begin{align}
  g^*(x)T_+f(x)&=V(x)\,\phi_0(x)^2
  Q\bigl(\eta(x)\bigr)P\bigl(\eta(x-i\gamma)\bigr)=F(x),
  \label{F(x)}\\
  g^*(x)T_-f(x)&=V^*(x)\phi_0(x)^2
  Q\bigl(\eta(x)\bigr)P\bigl(\eta(x+i\gamma)\bigr)=G(x),
  \label{G(x)}\\
  (T_+g)^*(x)f(x)&=V(x+i\gamma)\phi_0(x+i\gamma)^2
  Q\bigl(\eta(x+i\gamma)\bigr)P\bigl(\eta(x)\bigr)=F(x+i\gamma),
  \label{F(x+igamma)}\\
  (T_-g)^*(x)f(x)&=V^*(x-i\gamma)\,\phi_0(x-i\gamma)^2
  Q\bigl(\eta(x-i\gamma)\bigr)P\bigl(\eta(x)\bigr)=G(x-i\gamma).
  \label{G(x-igamma)}
\end{align}
Therefore the necessary and sufficient condition for the hermiticity
of the Hamiltonian becomes
\begin{equation}
  \int_{x_1}^{x_2}\bigl(F(x)+G(x)\bigr)dx
  =\int_{x_1}^{x_2}\bigl(F(x+i\gamma)+G(x-i\gamma)\bigr)dx.
  \label{intF+G}
\end{equation}
Of course it is required that there is no singularity on the integration
contours.

Let $C_{\pm}$ be the rectangular contours
$x_1\to x_2\to x_2\pm i\gamma\to x_1\pm i\gamma\to x_1$ and $D_{\pm}$
be the regions surrounded by $C_{\pm}$ including the contours.
Under the assumption that $F(x)$ and $G(x)$ do not have singularities on
$C_+$ and $C_-$ respectively\footnote{
If there are singularities on the contours $x_2\to x_2\pm i\gamma$ or
$x_1\pm i\gamma\to x_1$, we deform the contours and redefine $C_{\pm}$
and $D_{\pm}$ in order to avoid singularities on $C_{\pm}$.
For simplicity we have assumed no singularity on $C_{\pm}$ in the text.
}, 
the residue theorem implies that \eqref{intF+G} is rewritten as
\begin{align}
  &\int_0^{\gamma}\bigl(F(x_2+iy)-F(x_1+iy)-G(x_2-iy)+G(x_1-iy)\bigr)dy\n
  &\quad=2\pi\frac{\gamma}{|\gamma|}\Bigl(
  \sum_{\text{$x$\,:\,pole\,\,in\,$D_+$}}\!\!\!\text{Res}_xF(x)
  -\!\!\!\sum_{\text{$x$\,:\,pole\,\,in\,$D_-$}}\!\!\!\text{Res}_xG(x)\Bigr).
  \label{intF+G=res}
\end{align}

We will mention several sufficient conditions for \eqref{intF+G=res}.
If $F(x)$ and $G(x)$ are holomorphic in $D_+$ and $D_-$ respectively,
the r.h.s. of \eqref{intF+G=res} vanishes.
In the following we assume this.

\noindent
\underline{case 1}: $x_1=-\infty$, $x_2=\infty$.\\
If $\phi_0(x)$ is rapidly  decreasing (e.g. exponential in $\eta(x)$)
at $x\sim\pm\infty$, then \eqref{intF+G=res} is satisfied.

\noindent
\underline{case 2}: $x_1=0$, $x_2=\infty$.\\
If $\phi_0(x)$ is rapidly decreasing (e.g. exponential in $\eta(x)$)
at $x\sim\infty$, then \eqref{intF+G=res} becomes
$\int_0^{\gamma}\bigl(F(iy)-G(-iy)\bigr)dy=0$.
This is satisfied if $F(iy)=G(-iy)$, $y\in(0,\gamma)$.
As a sufficient condition for $F(iy)=G(-iy)$, we give the following
three reflection properties:
\begin{equation}
  \phi_0(-x)=\phi_0(x),\quad \eta(-x)=\eta(x),\quad V^*(x)=V(-x).
\end{equation}

\noindent
\underline{case 3}: $x_2=x_1+\omega$ ($0<\omega<\infty$).\\
The condition \eqref{intF+G=res} becomes
$\int_0^{\gamma}\bigl(F(x_1+\omega+iy)-F(x_1+iy)-G(x_1+\omega-iy)
+G(x_1-iy)\bigr)dy=0$.
This is satisfied if $F(x_1+\omega+iy)=G(x_1+\omega-iy)$
and $F(x_1+iy)=G(x_1-iy)$, $y\in(0,\gamma)$.
As a sufficient condition for $F(x_1+\omega+iy)=G(x_1+\omega-iy)$
and $F(x_1+iy)=G(x_1-iy)$, we give the following reflection relations
and the periodicity:
\begin{align}
  &\phi_0(-x+x_1)=\phi_0(x+x_1),\quad \eta(-x+x_1)=\eta(x+x_1),\quad
  V^*(x+x_1)=V(-x+x_1),\n
  &\phi_0(x+2\omega)=\phi_0(x),\quad \eta(x+2\omega)=\eta(x),\quad
  V(x+2\omega)=V(x).
\end{align}

\section{Eigenvalues and eigenvectors for an upper triangular matrix}
\label{app:upper}
\setcounter{equation}{0}
\renewcommand{\theequation}{C.\arabic{equation}}

Let $A=(a_{ij})_{1\leq i,j\leq n}$ be an upper triangular matrix,
namely $a_{ij}=0$ for $i>j$.
Its eigenvalue are $\alpha_i=a_{ii}$ ($i=1,\ldots,n$).
When $\alpha_i$'s are mutually distinct, the eigenvector corresponding to
the eigenvalue $\alpha_i$ is given by the determinant
\begin{equation}
  \begin{vmatrix}
    \bm{e}_1&\bm{e}_2&\bm{e}_3&\cdots&\bm{e}_i\\
    \alpha_1-\alpha_i&a_{12}&a_{13}&\cdots&a_{1i}\\
    &\alpha_2-\alpha_i&a_{23}&\cdots&a_{2i}\\
    &&\ddots&\ddots&\vdots\\
    \text{\LARGE $0$}&&&\alpha_{i-1}-\alpha_i&a_{i-1\,i}
  \end{vmatrix}\,,
  \label{eigenvec}
\end{equation}
where $\{\bm{e}_i\}$ is the natural basis, $(\bm{e}_i)_j=\delta_{ij}$.


\end{document}